\documentclass[preprint,5p,times,twocolumn]{elsarticle}

\usepackage{amssymb}
\usepackage{amsmath}
\usepackage[utf8]{inputenc} 
\usepackage[T1]{fontenc}    
\usepackage{hyperref}       
\usepackage{xurl}
\usepackage{url}            
\usepackage{booktabs}      
\usepackage{amsfonts}       
\usepackage{nicefrac}       
\usepackage{tikz}
\usepackage{multicol}
\usepackage{multirow}
\usepackage{mathtools}
\usepackage{pifont}
\usepackage{pgfplots}
\usetikzlibrary{shapes,arrows}
\pgfplotsset{compat=1.16}
\usepackage{array}
\usepackage{enumitem}
\usepackage{lscape}
\usepackage{longtable}
\usepackage{tcolorbox}
\usepackage{subcaption}
\usepackage{wrapfig}
\usepackage{cellspace}
\usepackage{microtype}
\usepackage{lipsum}
\usepackage{graphicx}

\usetikzlibrary{calc} 
\usetikzlibrary{positioning}
\definecolor{darkblue}{RGB}{95,138,168}
\definecolor{lightblue}{RGB}{95, 122, 150}

\journal{journal}
\begin{document}
\begin{frontmatter}

\title{Sustainable business decision modelling with blockchain and digital twins: A survey}

\author{Gyan Wickremasinghe, Siofra Frost, Karen Rafferty, Vishal Sharma}

\affiliation{organization={School of Electronics, Electrical Engineering and Computer Science (EEECS) \\Queen's University Belfast},
            country={Northern Ireland, United Kingdom}}

%% Abstract
\begin{abstract}
Industry 4.0 and beyond will rely heavily on sustainable Business Decision Modelling (BDM) that can be accelerated by blockchain and Digital Twin (DT) solutions. BDM is built on models and frameworks refined by key identification factors, data analysis, and mathematical or computational aspects applicable to complex business scenarios. Gaining actionable intelligence from collected data for BDM requires a carefully considered infrastructure to ensure data transparency, security, accessibility and sustainability. Organisations should consider social, economic and environmental factors (based on the triple bottom line approach) to ensure sustainability when integrating such an infrastructure. These sustainability features directly impact BDM concerning resource optimisation, stakeholder engagement, regulatory compliance and environmental impacts. To further understand these segments, taxonomies are defined to evaluate blockchain and DT sustainability features based on an in-depth review of the current state-of-the-art research. Detailed comparative evaluations provide insight into the reachability of the sustainable solution in terms of ideologies, access control and performance overheads. Several research questions are put forward to motivate further research that significantly impacts BDM. Finally, a case study based on an exemplary supply chain management system is presented to show the interoperability of blockchain and DT with BDM.
\end{abstract}

%% Keywords

\end{frontmatter}

\section{Introduction}
\label{sec1}
Businesses and organisations are urged to adopt more sustainable and eco-friendly practices in the face of increasing environmental degradation, resource depletion and raw material scarcity~\cite{GUPTA2017242}. This urgency is heightened by rapid industrialisation, population growth and fierce commercial competition ~\cite{GUPTA2017242}. In this context, Sustainable Supply Chain Management (SSCM) is an effective tool for improving efficiency~\cite{KOC2023108820, GUPTA2017242}.

However, avoiding negative social and environmental consequences while improving sustainability remains difficult~\cite{KOBERG20191084}. In this space, Business Decision Modelling (BDM) is an essential practice for SSCM, which enables organisations to reduce the risk of poor decisions throughout business processes by utilising modelling and simulation technologies~\cite{RAMADAN7093873, VERHAGEN9321718}. Furthermore, as the supply chain has globalised, the demand to help monitor and access Supply Chain Management Systems (SCMS) has increased for mitigating risks and disruptions~\cite{NGUYEN2022108381}. Amid these concerns and rapid economic growth, the digitisation of the supply chain, known as Industry 4.0, holds significant potential for further discussions and improvements in sustainability within BDM~\cite{BALASUBRAMANIAN9566480, LUTHRA2022102582, SRHIR2023138111}.

\begin{table*}[ht]
\tiny
  \centering
  \resizebox{15cm}{!}{
  \begin{tabular}{|c|c|c|c|c|c|c|}
    \hline
    \textbf{Technology} & \textbf{Features} & \multicolumn{2}{c|}{\textbf{Social SDGs}} & \multicolumn{2}{c|}{\textbf{Environmental SDGs}} & \textbf{Economic SDGs} \\
    \cline{3-7}
     &  & \textbf{S1} & \textbf{S2} & \textbf{Ev1} & \textbf{Ev2} & \textbf{Ec1} \\
    \hline
    \multirow{5}{*}{{Blockchain} \cite{LUND8823906,LENG2020110112,REJEB2023200126}}  
     & {Consensus Protocol} & {\ding{51}} & {\ding{55}} & {\ding{51}} & {\ding{55}} & {\ding{51}}\\
     & {Smart Contract} & {\ding{55}} & {\ding{55}} & {\ding{51}} & {\ding{55}} & {\ding{51}}\\
     & {Security \& Privacy} & {\ding{51}} & {\ding{51}} & {\ding{55}} & {\ding{55}} & {\ding{51}}\\
     & {Scalability} & {\ding{55}} & {\ding{51}} & {\ding{51}} & {\ding{51}} & {\ding{55}}\\
     & {Storage Efficiency} & {\ding{55}} & {\ding{51}} & {\ding{51}} & {\ding{51}} & {\ding{55}}\\
    \hline
    \multirow{3}{*}{{Digital Twin} \cite{NGUYEN2022108381,SINGH2023109172, KAMBLE2022121448}} 
     & {Data Integration} & {\ding{55}} & {\ding{51}} & {\ding{51}} & {\ding{51}} & {\ding{51}}\\
     & {Predictive Analysis} & {\ding{55}} & {\ding{51}} & {\ding{51}} & {\ding{51}} & {\ding{51}}\\
     & {Monitoring} & {\ding{55}} & {\ding{51}} & {\ding{55}} & {\ding{55}} & {\ding{51}}\\
    \hline
  \end{tabular}
  }
  \caption{Features of blockchain and DT impacted by the SDGs (S1 - Reduced Inequalities; S2 - Sustainable Cities and Communities; Ev1 - Affordable and Cleaner Energy; Ev2 - Climate Action; Ec1 - Industry, Innovation and Infrastructure).}
  \label{tab: SDG}
\end{table*}

BDM leverages digitisation to significantly boost business value through effective knowledge management~\cite{KOC2023108820}. To develop an efficient global supply chain securely and transparently, emerging technologies such as blockchain - a decentralised distributed ledger - play a role, particularly within the consensus layer~\cite{Silva_2024, WANG8629877}. This layer is responsible for maintaining integrity across the blockchain by ensuring all nodes agree on the same state, as it aims to balance minimising complexity with maximising transactions per second (TPS)~\cite{WANG8629877, XIAO8972381}. Concurrently, to improve decision-making models within BDM, a Digital Twin (DT) can be utilised to offer space in which a simulation model of a Physical Twin's (PT) virtual counterpart can be run. DT can use historical data to enable scenario manipulation and analysis, further improving the understanding of Supply Chain Management (SCM)~\cite{SUHAIL10.1145/3517189, DIETZ8966454, TAO8477101}.

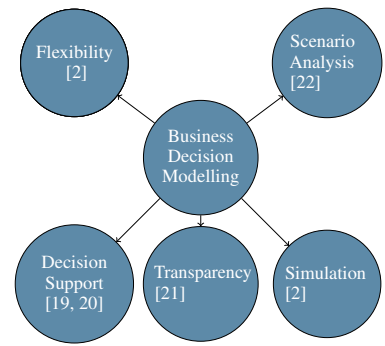
\begin{figure}
    \centering
    \scalebox{0.75}{
    \begin{tikzpicture}[scale=1.5]
    
    \node[draw, circle, fill=darkblue, text=white] (business) at (0,0) {\parbox{2.2cm}{\centering Business Decision\\ Modelling}};
    
    \node[draw, circle, fill=lightblue, text=white] (decision_support) at (-2,-1.5) {\parbox{1.8cm}{\centering Decision\\ Support \\\cite{KIR2021101639,KELLNER2019505}}};
    \node[draw, circle, fill=lightblue, text=white] (flexibility) at (-1.5,1.5) {\parbox{1.8cm}{\centering Flexibility\\~\cite{KOC2023108820}}};
    \node[draw, circle, fill=lightblue, text=white] (transparency) at (0,-2) {\parbox{1.8cm}{\centering Transparency\\~\cite{CENTOBELLI2022103508}}};
    \node[draw, circle, fill=lightblue, text=white] (scanrio_analysis) at (1.5,1.5) {\parbox{1.8cm}{\centering Scenario\\ Analysis\\~\cite{yan_integrated_2022}}};
    \node[draw, circle, fill=lightblue, text=white] (simulation) at (2,-1.5) {\parbox{1.8cm}{\centering Simulation\\~\cite{KOC2023108820}}};
    
    \draw[->] (business) -- (decision_support);
    \draw[->] (business) -- (flexibility);
    \draw[->] (business) -- (transparency);
    \draw[->] (business) -- (scanrio_analysis);
    \draw[->] (business) -- (simulation);
    
    \end{tikzpicture}}
    \vspace{2\baselineskip}
    \caption{Features impacting sustainability in BDM.}
    \label{fig: BDM features}
\end{figure}

Additionally, these advancements support the 17 Sustainable Development Goals (SDGs) established by the United Nations (UN), which encourage organisations and countries to prioritise sustainability~\cite{GUPTA2017242}. The growing trend of disclosing Environmental, Social, and Governance (ESG) goals aligns closely with the SDGs. As governments push for transparency, organisations are increasingly reporting on their ESG initiatives, which contribute directly to SDG targets such as responsible consumption, climate action, decent work and economic growth. This proactive approach not only strengthens stakeholder trust and attracts sustainable investment but also supports regulatory alignment and enhances a company’s reputation and operational resilience, positioning it for long-term impact aligned with global sustainability objectives~\cite{wilburn2020esg, ISIK2024105114}. As a result, these goals influence the technologies used in supply chains. The interaction of blockchain and DT could be of potential use in Industry 4.0, with Table~\ref{tab: SDG} identifying the impact of their features on various SDGs~\cite{SUHAIL10.1145/3517189}. Despite these advances, there is still untapped potential in exploring sustainability within BDM using blockchain and DT, emphasising the importance of additional research in this field~\cite{TSAO2021110452, yan_integrated_2022}.
\newline
The following section explores the application of BDM through its architecture, application, and challenges, along with a comprehensive overview of its continued presence in the digital age.

\subsection{BDM}
BDM offers a structured approach that involves applying logic-based or mathematical models to aid in efficient decision-making and can severely impact various sustainability features, as highlighted in Fig.~\ref {fig: BDM features}. BDM can have extensive involvement in decision-making, particularly in SCM. It, therefore, plays a critical role in guiding organisations toward informed, effective and efficient business decisions, risk assessment and process optimisation~\cite{KOC2023108820}.

\begin{figure}[hb]
  \centering
  \includegraphics[width=0.5\textwidth]{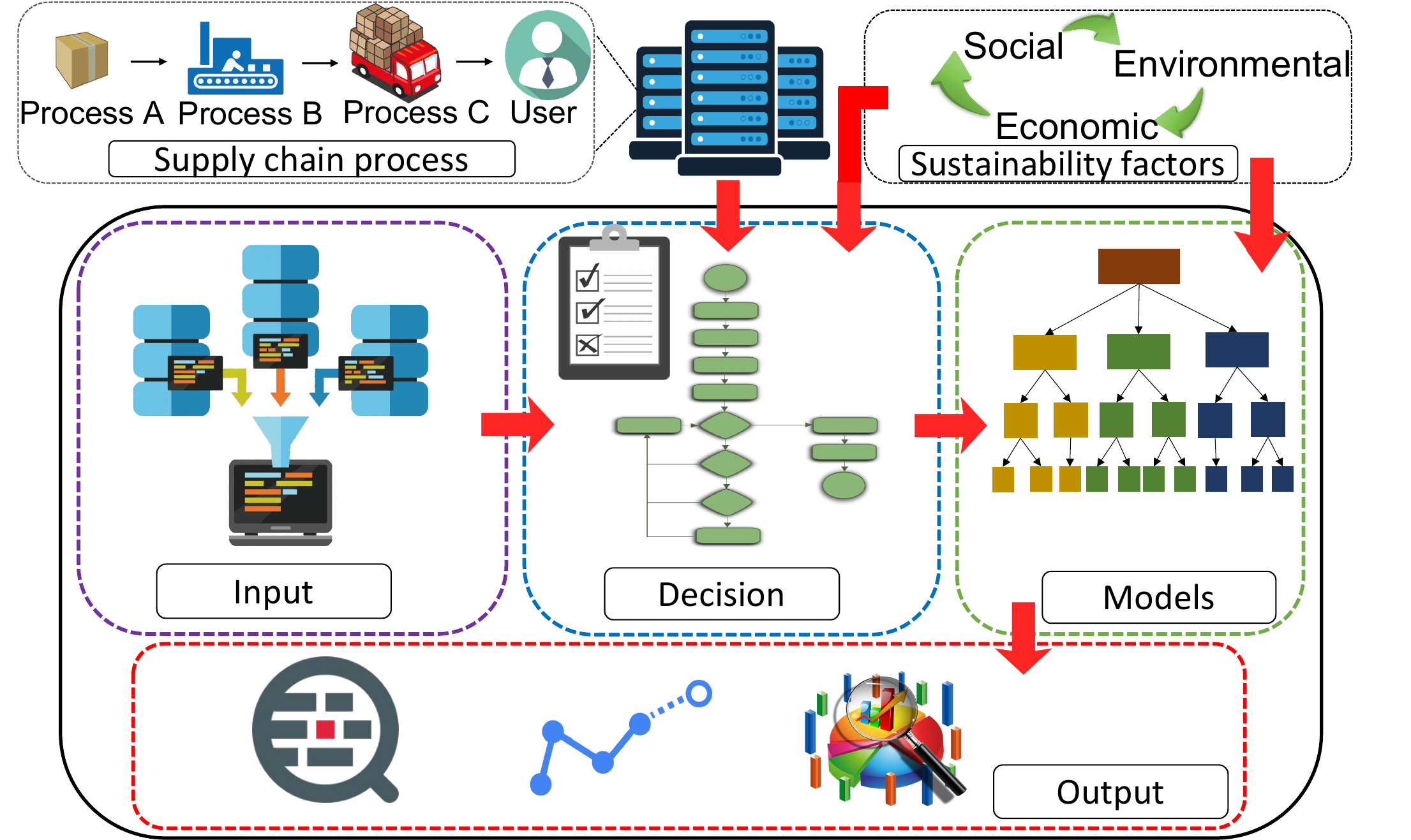}
  \caption{An Illustration of the Exemplary Workflow of BDM.}
  \label{fig: BDM architecture}
\end{figure}

\begin{figure}[!ht]
  \centering
  \includegraphics[width=0.5\textwidth]{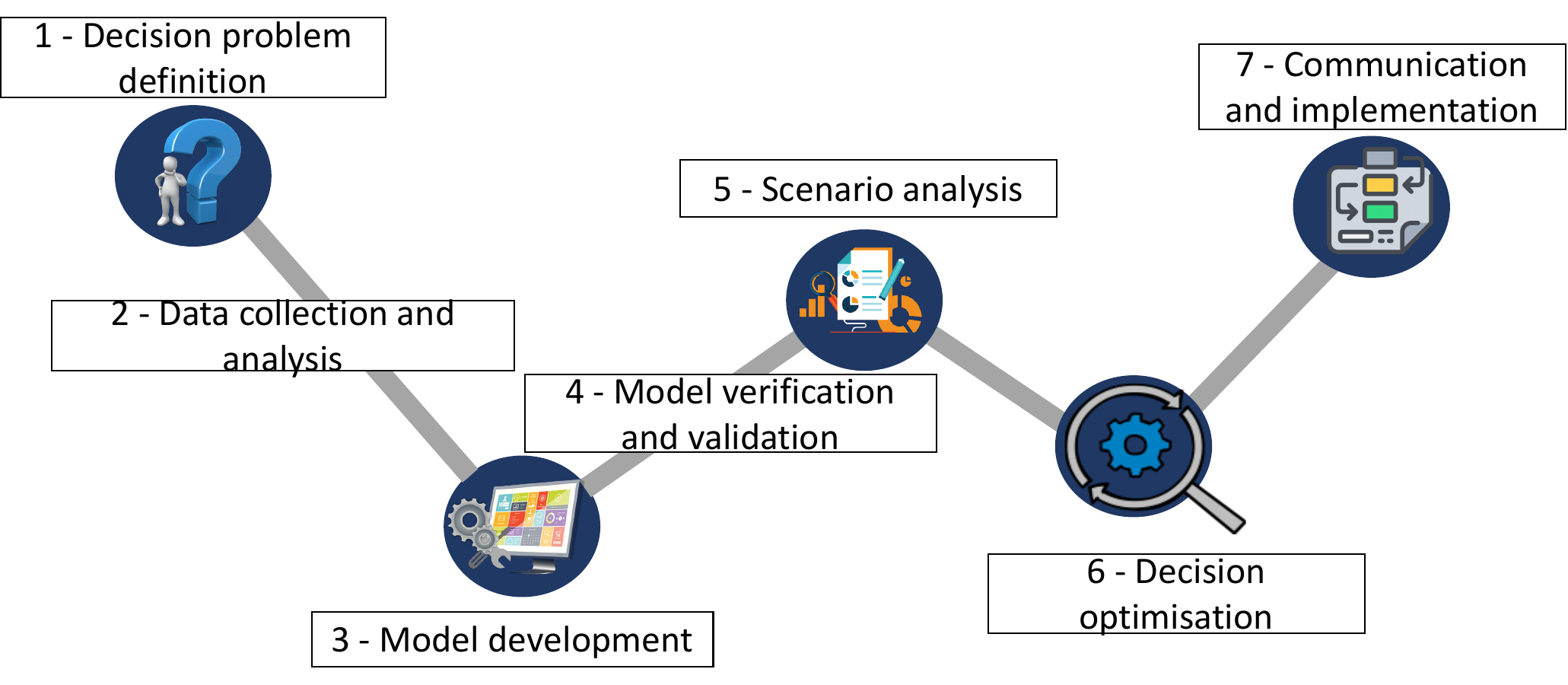}
  \caption{An illustration of the 7-step BDM process.}
  \label{fig: BDM process}
\end{figure}

\subsubsection{Architecture}

BDM's architectural design follows a four-step workflow, as shown in Fig.~\ref{fig: BDM architecture}. The procedure begins with the user entering data into the system, which includes parameters ranging from customer requirements to supplier capabilities. Decision-makers then weigh these inputs against various criteria, such as cost, quality and sustainability, to determine the best action. BDM strategies illustrate the selected decisions in the next step of the workflow. The decision trees can be utilised for further improvement via an optimisation model in this domain. This allows the manipulation of multiple inputs and produces a variety of outputs, such as production schedules, inventory levels and supplier selections. The final step involves the simulation of a supply chain management system to provide predictive insights and enable informed changes. 

\subsubsection{Applications}

BDM is a versatile process (exemplary BDM processes are depicted in Fig.~\ref{fig: BDM process}) with numerous applications, notably in SCM, which includes aids in cash flow forecasting, budgeting and investment analysis via financial modelling. It can also help with risk management, operational management (including production planning, scheduling and quality control) and marketing strategies such as product pricing and profitability. Understandably, much modelling research in BDM focuses on applications significantly impacting SCM. For example, Koc \textit{et al.}~\cite{KOC2023108820} emphasised the importance of sustainable supplier selection by employing decision-making models. 

\begin{figure}[!h]
  \centering
  \includegraphics[width=0.45\textwidth]{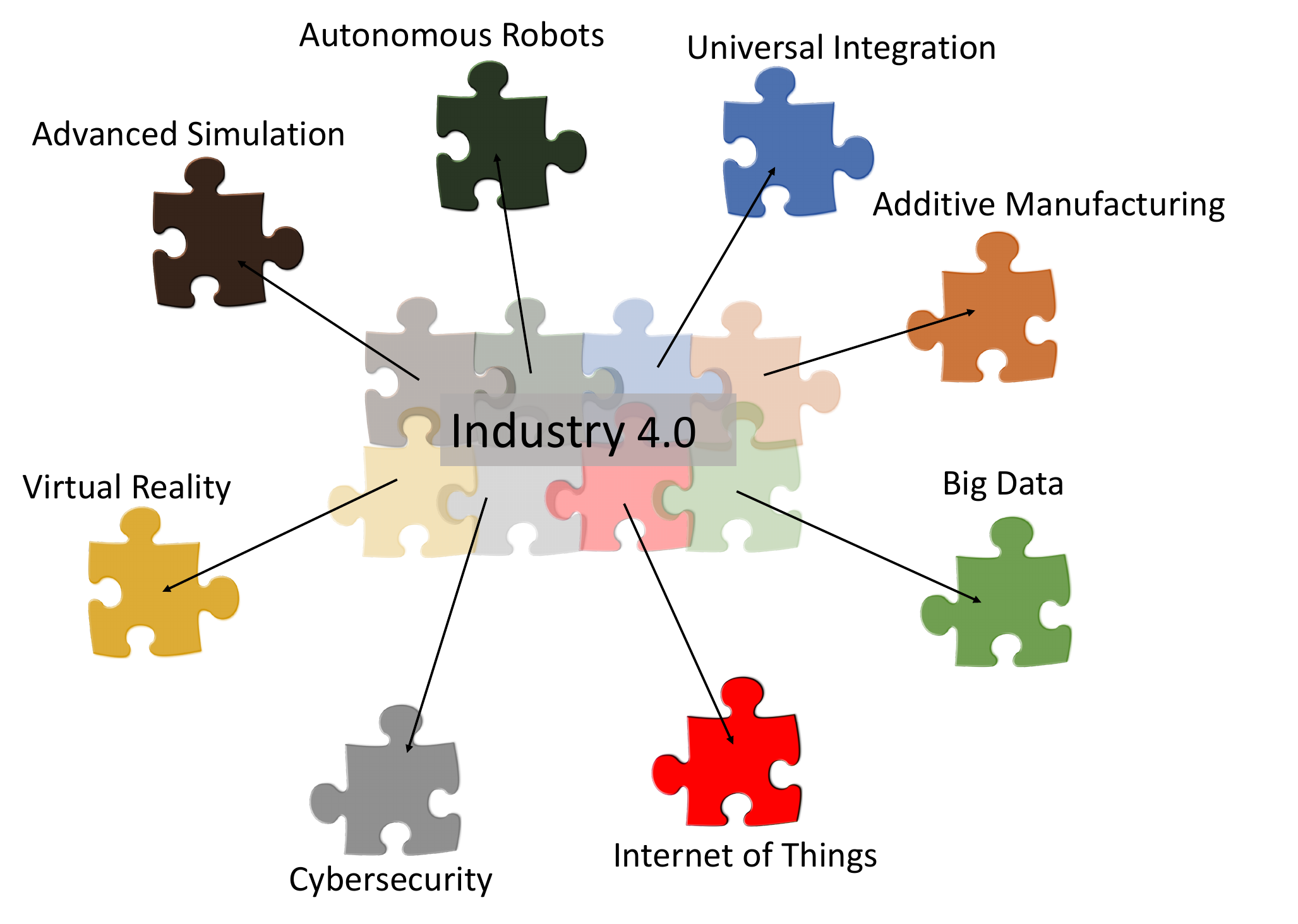}
  \caption{The Pillars of Industry 4.0.}
  \label{fig: Industry 4.0}
  \end{figure}
\begin{figure}[!h]
  \centering
  \includegraphics[width=0.4\textwidth]{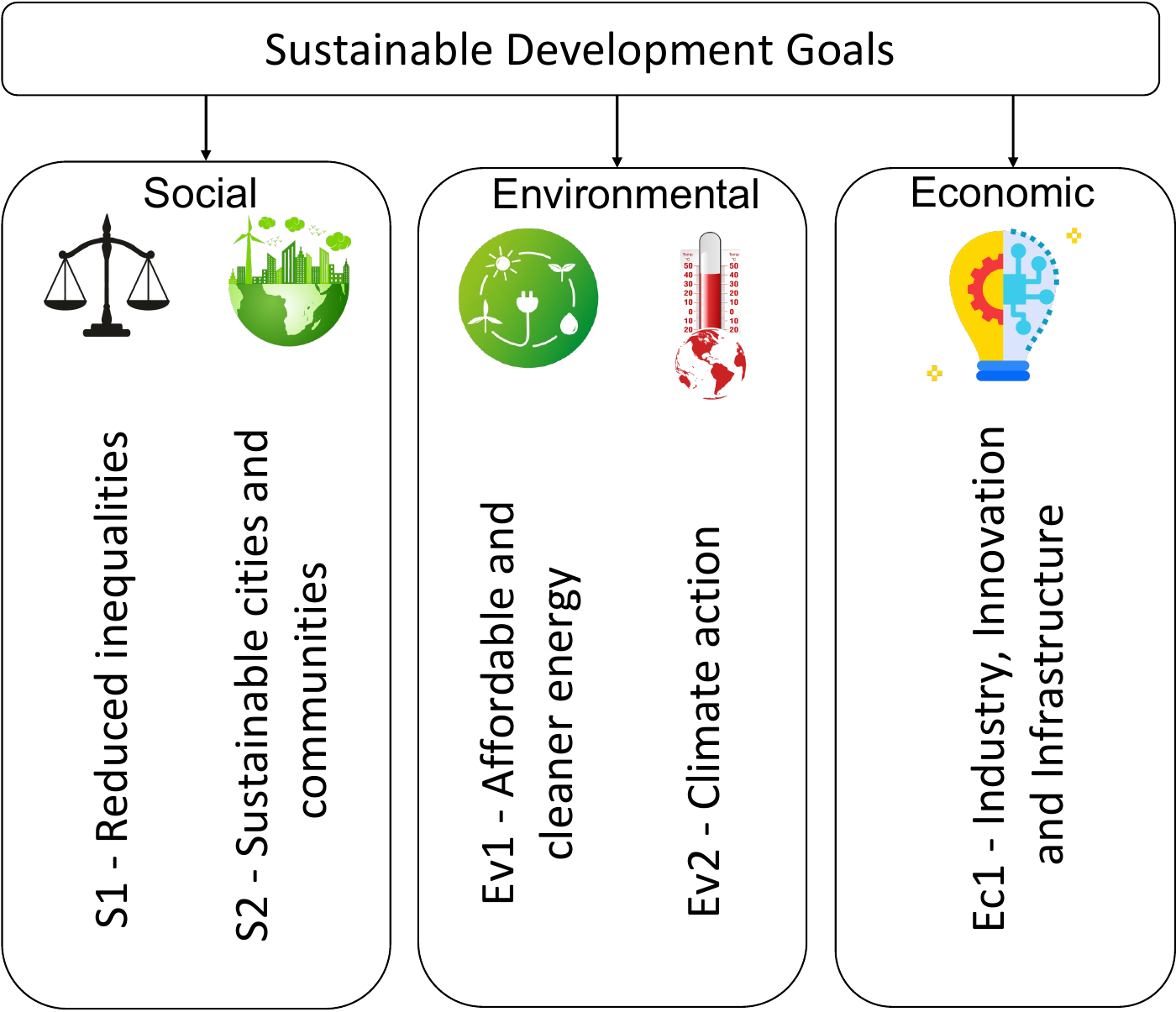}
  \caption{Sustainable Development Goals impacting Industry 4.0.}
  \label{fig: SDGs}
\end{figure}

\subsubsection{Challenges}

With the advent of Industry 4.0 (see Fig.~\ref{fig: Industry 4.0}), BDM faces many new challenges. Typically, decision models face difficulties expressing conditional independencies, managing an exponential increase in size with the number of variables and requiring probability preprocessing~\cite{DIEZ20181}. Furthermore, building multi-level and collaborative models is a challenge. Data quality and availability are also significant issues, especially when data formats differ, and data extraction from multiple sources becomes complex. Challenges around limited transparency and accountability exacerbate these issues, while cost considerations emerge as critical constraints that must be carefully managed within BDM~\cite{MONCH2011557}.

\subsection{Sustainability in BDM}

%  Revised Text
Incorporating the UN's SDGs into decision-making processes drives sustainability in BDM, as illustrated in  Fig.~\ref{fig: SDGs}. The overarching goal of sustainability in BDM is to ensure that the long-term consequences of organisational actions on the environment, society and economy are considered in decision-making. This necessitates a comprehensive strategy considering every stage of a product's or service's lifecycle $-$ ranging from raw material extraction to eventual disposal. Another aspect of sustainable BDM is engaging stakeholders such as consumers, employees and local communities to ensure their perspectives and concerns in decision-making~\cite{GUPTA2017242, KOBERG20191084}.

\subsubsection{Architecture}

The architecture of sustainable BDM is similar to that of traditional BDM structures. It further incorporates sustainability features into the workflow. This implies that the data simulates the supply chain's response under varying sustainability conditions during the decision-making and modelling phases~\cite{KOC2023108820}.

\subsubsection{Applications}

% Revised text
Sustainable BDM aims to improve an organisation's environmental, economic and social performance~\cite{KOBERG20191084}. This can be accomplished by incorporating sustainability criteria into selecting suppliers, investment decisions and product design processes. Furthermore, sustainable BDM can reveal resource efficiency and waste reduction opportunities, resulting in significant cost savings for businesses.

\subsubsection{Challenges}

Sustainability in BDM will impact the supply chain structure. This is a significant challenge as it requires stakeholder engagement and expertise in environmental, economic, and social sciences. There is uncertainty around the long-term impacts as decision outcomes would vary, and it becomes more complex when including social, economic and environmental factors in a business process. Furthermore, stakeholders may resist changes due to concerns about data access and transparency~\cite{KARMAKER2023108806}.

\subsection{Blockchain in BDM}

With the globalisation of the supply chain and issues around a single point of failure in centralised databases, blockchain in BDM aims to resolve these problems through the immutable and transparent nature of the technology \cite{KOBERG20191084, DEVILLIERS2021598}.

\subsubsection{Architecture}
\begin{figure}
  \centering
  \includegraphics[width=0.5\textwidth]{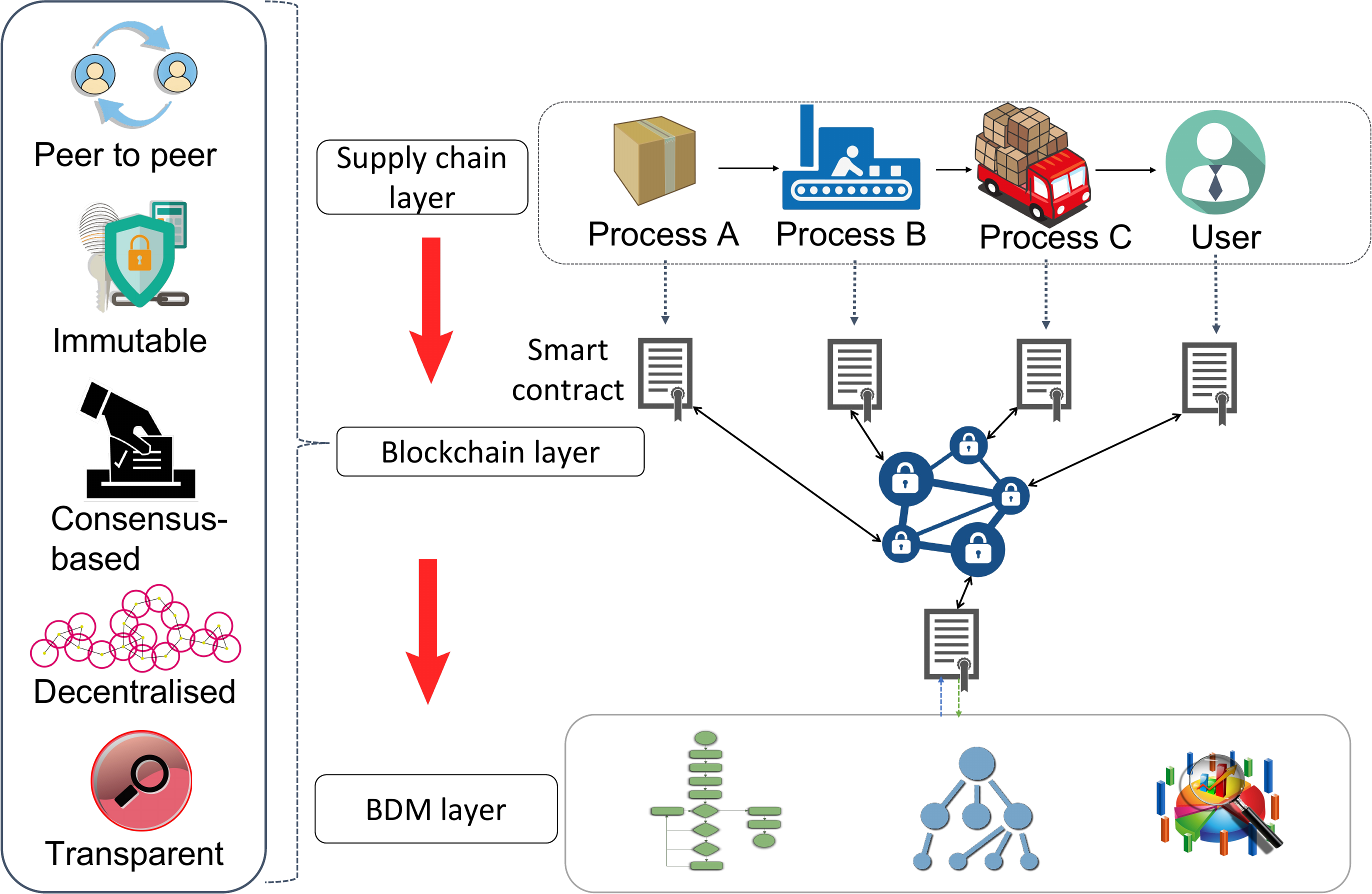}
  \caption{A workflow of the blockchain-based BDM.}
  \label{fig: Blockchain BDM architecture}
\end{figure}

Blockchain can be integrated as a component of BDM, and its decentralised architecture replaces the prevalent centralised data storage systems. The blockchain architecture follows five layers (refer to Section \ref{sec: Section 4}). The consensus layer facilitates decentralisation within the peer-to-peer network. The architecture of a blockchain-based BDM is depicted in Fig. ~\ref{fig: Blockchain BDM architecture}, in which each node in the blockchain maintains a copy of the supply chain data. The architecture can be used to implement smart contracts, which are self-executing and programmable agreements that serve as automated decision-making tools based on predefined rules and conditions.

\subsubsection{Applications}

Extensive research has been conducted into blockchain applications and their interoperability with technologies such as the Internet of Things, Edge computing and DT. Current blockchain-based BDM applications include SCM, healthcare, governance and financial management, where continued growth is indicating the further potential for this technology~\cite{WANG10.1145/3582882, BELCHIOR10.1145/3471140, HUANG10.1145/3441692}. Blockchain could improve data accuracy, decision-making processes, transparency and stakeholder trust. Furthermore, introducing smart contracts broadens the utility of blockchain-based BDM beyond traditional domains~\cite{REJEB2023200126, HUANG10.1145/3441692, YOUSEFI2024110577}.

\subsubsection{Challenges}

Transitioning from a centralised database to a blockchain presents challenges with integration, scalability, regulations and compliance complexities~\cite{FAHMIDEH10.1145/3530813}. Within BDM, a significant challenge is the efficient management of data affected by the constant exchange between the BDM and supply chain layer, as seen in Fig. \ref{fig: Blockchain BDM architecture}. For example, constantly appending BDM processes to the blockchain can affect latency by block mining which affects the throughput and performance~\cite{YE10004754,LENG2020110112}.

\subsection{DT in BDM}

A DT is a dynamic simulation model that creates a virtual replica of a physical entity based on real-time or on-demand analysis. This replica, which is constantly updated, can simulate the behaviour of physical systems. The applications of DT include preventive maintenance, testing alterations to the physical system and other tasks that would yield high capital expenditure (CAPEX)/operating expenses (OPEX) or incur risk when configuring the actual system.

\subsubsection{Architecture}
\begin{figure}
  \centering
  \includegraphics[width=0.5\textwidth]{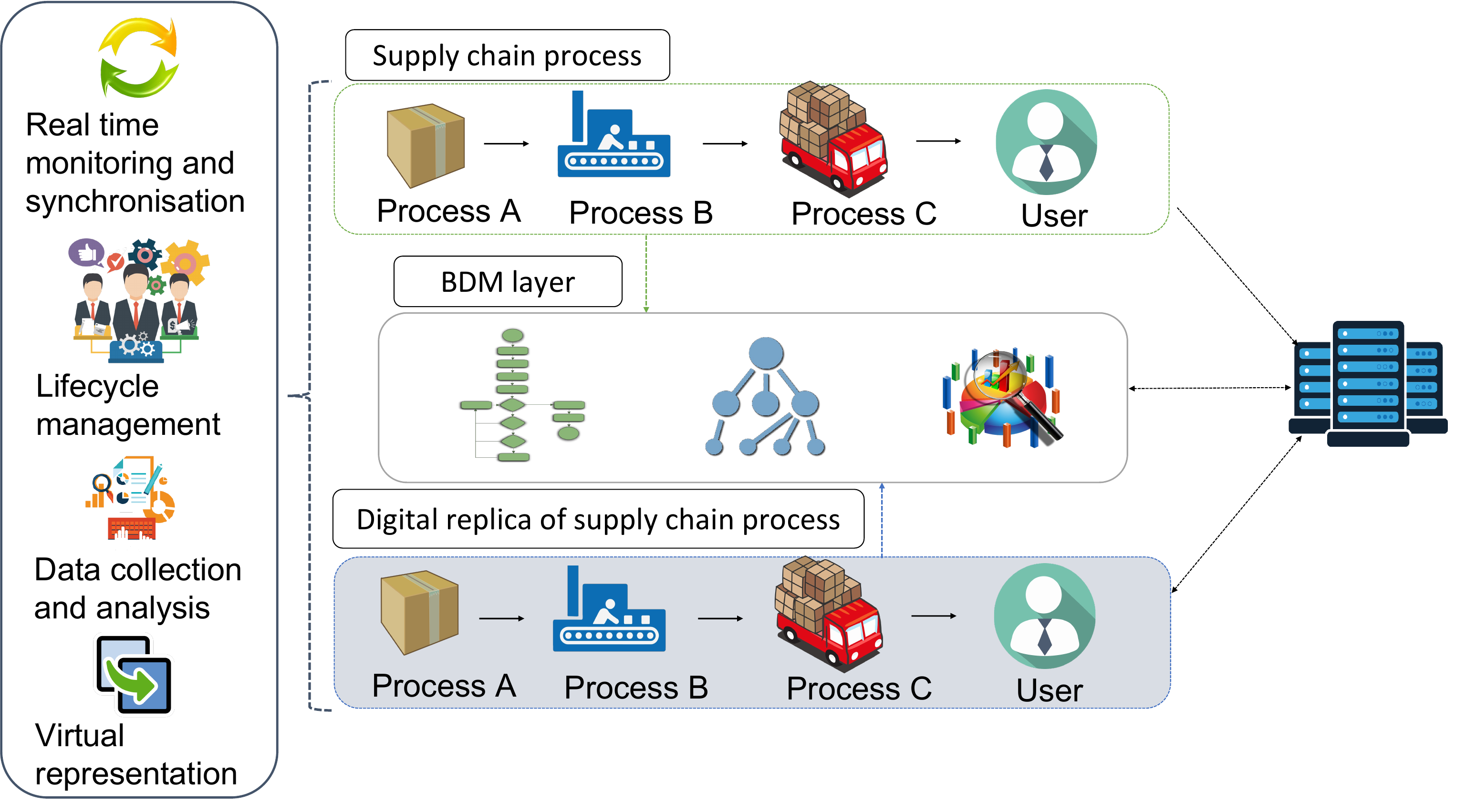}
  \caption{An exemplary workflow model of DT-based BDM.}
  \label{fig: DT BDM architecture}
\end{figure}

The three main layers of DT-based BDM architecture are depicted in Fig.~\ref{fig: DT BDM architecture}. The physical supply chain is the first layer, which includes the assets, devices, sensors and equipment that generate data and interact with the real world. The digital supply chain, the second layer, is a real-time virtual replica of the physical supply chain. This enables real-time data simulation throughout the decision-making and modelling stages. The final data layer, the BDM layer, is a database that stores all data related to the physical space and DTs.

\subsubsection{Applications}
From the supply chain perspective, integrating DT in BDM can improve modelling of the flow of goods, information and finances. This can help optimise processes and resources, detect bottlenecks and increase stakeholder engagement~\cite{NGUYEN2022108381, MANDOLLA2019134}. Another critical application is predictive maintenance, in which DT creates virtual representations of infrastructure for real-time health monitoring, failure prediction and maintenance activity of the assets.

\subsubsection{Challenges} 

DT-based BDM relies heavily on reliable data, which poses a significant challenge as data inaccuracies and inconsistencies can negatively affect simulation models, resulting in additional costs to the organisation. Another barrier is the lack of standard protocols for creating and deploying DTs, which increases the complexity of their integration into existing systems. Furthermore, DT-based BDM generates large models that consume a significant amount of energy and generate large amounts of data, posing another challenge to be addressed before practical deployment~\cite{DIETZ8966454, KAMBLE2022121448}.

\textbf{Structure of the article:} The remaining parts of this article are organised as follows. Section \ref{sec: Section 2} explores the different surveys and reviews related to SSCM that identify blockchain or DT to support BDM. Section \ref{sec: Section 3} discusses the factors affecting sustainability in BDM based on the triple bottom line approach, looking at social, economic and environmental properties. Section \ref{sec: Section 4} identifies the different sustainability factors within a blockchain and Section \ref{sec: Section 5} identifies these factors within DT technology. Section \ref{sec: Section 6} explores the existing computing problems in relation to BDM, blockchain and DT and identifies the potential research trends. Section \ref{sec: Section 7} presents a case study to evaluate the technological benefits of sustainability factors in BDM. Section \ref{sec: Section 8} defines the research methodology. Finally, the article is concluded in Section \ref{sec: Section 9}.

\section{Comparison with other surveys}\label{sec: Section 2}

\begin{table*}[!ht]
\tiny
    \centering
    \resizebox{18cm}{!}{
    \begin{tabular}{|c|p{5cm}|p{2.3cm}|c|c|c|}
        \hline
        \multirow{2}{*}{\textbf{Article}} & \multirow{2}{*}{\textbf{Research focus}} & \multirow{2}{*}{\textbf{Methodology}} & \multicolumn{2}{c|}{\textbf{SSCM}} & \multirow{2}{*}{\textbf{BDM}} \\
        \cline{4-5}
        & & & \textbf{Blockchain} & \textbf{DT} & \\
        \hline
        %2023 - Nov
        Adu-Amankwa \textit{et al.}~\cite{ADUAMANKWA2023105064} & Impact of blockchain and DT for building life cycle management & Systematic Review & {\ding{51}} & {\ding{51}} & {\ding{55}}
        \\
        \hline
        %2024 - July
        Es-haghi \textit{et al.}~\cite{ESHAGHI2024107342} & Identifying tools and methods that govern a DT to reduce computation demand of DTs in a cyber-physical system & Survey & {\ding{55}} & {\ding{51}} & {\ding{55}}
        \\
        \hline
        %2020 August
        Esmaeilian \textit{et al.}~\cite{ESMAEILIAN2020105064} & Industry 4.0 and blockchain towards a sustainable supply chain by business models & Survey & {\ding{51}} & {\ding{55}} & {\ding{51}}\\
        \hline
        %2024 - Nov
        Figueiredo \textit{et al.}~\cite{FIGUEIREDO2024111018} & Explores the integration of blockchain and DT for Building engineering to support sustainability & Systematic Review & {\ding{51}} & {\ding{51}} & {\ding{55}}
        \\
        \hline
        %2024 - May
        Ghasmei \textit{et al.}~\cite{GHASEMI2024100599} & How industry 4.0 can have an impact on system optimisation and production scheduling focusing on tackling current issues like resilience, production disruption and lean production & Systematic Review & {\ding{55}} & {\ding{51}} & {\ding{55}}
        \\
        \hline
        %2020 Dec
        Guo \textit{et al.}~\cite{GUO9279323} & The influence of blockchain with information disclosure within the fashion supply chain enabling sustainability & Survey & {\ding{51}} & {\ding{55}} & - \\ 
        \hline
        %2023 
        Hammi \textit{et al.}~\cite{HAMMI10.1145/3588999} & Security impacts of a digital supply chain & Survey & {\ding{51}} & {\ding{55}} & {\ding{55}}\\
        \hline
        %2024 - Oct
        Hafeez \textit{et al.}~\cite{HAFEEZ2024107796} & Evaluates impact of DTs on decarbonisation efforts in energy-intensive industries, with a focus on industrial furnace operations & Systematic Review & {\ding{55}} & {\ding{51}} & {\ding{55}}
        \\
        \hline
         %2021 Dec
        Kamble \textit{et al.}~\cite{KAMBLE2022121448} & Evaluation of Digital Supply Chain Twin architecture targeting sustainability & Systematic Review & {\ding{55}} & {\ding{51}} & {\ding{55}}\\
        \hline
        %2020 September
        Kopyto \textit{et al.}~\cite{KOPYTO2020120330} & Viewpoints by experts of the impact on blockchain towards SSCM & Empirical Review & {\ding{51}} & {\ding{55}} & {\ding{55}}\\
        \hline
        %2020 June
        Kouhizadeh \textit{et al.}~\cite{KOUHIZADEH2021107831} & Blockchain adaptation barriers to a sustainable supply chain & Bibliometric Review and Framework & {\ding{51}} & {\ding{55}} & {\ding{55}}\\
        \hline
        %2020 - Nov
        Lee \textit{et al.}~\cite{lee_intelligent_2020} & Methods of improving intelligent maintenance systems through industry 4.0 technologies & Survey & {\ding{51}} & {\ding{51}} & {\ding{55}}
        \\
        \hline
        %2023 - Oct
        Li \textit{et al.}~\cite{LI2023122794} & Barriers for organisations to adopt blockchain and DT for Industrial Internet of Things implementations, based on their decision-making approaches & Survey & {\ding{51}} & {\ding{51}} & -
        \\
        \hline
        %2024 - April
        Liu \textit{et al.}~\cite{LIU2024248} & Evaluation of blockchain-based federated learning for DT applications & Survey & {\ding{51}} & {\ding{51}} & {\ding{55}}
        \\
        \hline
        %2020 July
        Leng \textit{et al.}~\cite{LENG2020110112} & Methods in which blockchain can overcome potential barriers to achieving sustainability in manufacturing and product lifecycle management & Survey & {\ding{51}} & {\ding{55}} & {\ding{55}}\\
        \hline
        %2021 June
        Lohachab \textit{et al.}~\cite{LOHACHAB10.1145/3460287} & Blockchain interoperability and applications & Survey & {\ding{51}} & {\ding{55}} & {\ding{55}}\\
        \hline
        %2021 December
        Nguyen \textit{et al.}~\cite{NGUYEN2022108381} & Impact of DT and Physical Internet in Supply chain management & Systematic Review & {\ding{55}} & {\ding{51}} & {\ding{55}}\\
        \hline
        %2022 13 Dec
        Rejab \textit{et al.}~\cite{REJEB2023200126} & Identifying the impact of blockchain with the transition to a circular economy & Systematic Review & {\ding{51}} & {\ding{55}} & {\ding{55}}\\ 
        \hline
        %2022 8 September
        Sahoo \textit{et al.}~\cite{SAHOO9908297} & Supply chain visibility using blockchain & Systematic Review & {\ding{51}} & {\ding{55}} & - \\
        \hline
        %2022 - May
        Sahoo \textit{et al.}~\cite{SAHOO} & The application of blockchain for SSCM based on insights from certain sectors & Systematic Review & {\ding{51}} & {\ding{55}} & -
        \\
        \hline
        %2023 March
        Singh \textit{et al.}~\cite{SINGH2023109172} & Identify and analyse the role of DT technology in enhancing the resilience and sustainability of food supply chain & Bibliometric Review & {\ding{55}} & {\ding{51}} & {\ding{55}}\\
        \hline
        %2019 - Dec
        Weking \textit{et al.}~\cite{WEKING} & An empirical study which examines how blockchain impacts existing business models. Five business models were proposed: blockchain for business integration, multi-sided platforms, security, as an offering, and for monetary value & Survey and Empirical Review & {\ding{51}} & {\ding{55}} & -
        \\
        \hline
        %2022 12 September
        Yang \textit{et al.}~\cite{YANG9916158} & Evaluates the significance of emerging technology that enables supply chain resilience and sustainability & Bibliometric Review & {\ding{51}} & {\ding{55}} & {\ding{55}}\\
        \hline
        %2022 29 Dec
        Yontar~\cite{YONTAR2023117173} & Analyse and interpret the critical success factors with blockchain in the flow of agri-food supply chain management on the way to a circular economy & Survey & {\ding{51}} & {\ding{55}} & - \\
        \hline
        %2022 December
        Yousefi and Tosarkani~\cite{YOUSEFI10005278} & Impact of blockchain adoption on the improvement of SSCM & Bibliometric Review & {\ding{51}} & {\ding{55}} & {\ding{55}}\\
        \hline
        This survey & Achieving sustainability in BDM through enhancing sustainability practices in blockchain and DT & Survey & {\ding{51}} & {\ding{51}} & {\ding{51}}
        \\
        \hline
    \end{tabular}
    }
    \caption{Comparison with related survey articles ({\ding{51}} Discussed and specified; {\ding{55}} Not discussed and not specified; - Discussed or specified).}
    \label{table: Existing surveys}
\end{table*}

The number of review articles has grown in the last few years, as shown in Table~\ref{table: Existing surveys} and Fig.~\ref{fig: Published Years}. However, based on the context covered in existing articles, there is a research gap when considering sustainability in BDM using blockchain and DT. For example, blockchain presents security and transparency features, which benefit a DT with trusted data. This allows companies to test and validate new ideas and processes in a safe and controlled environment without the risk of disrupting or incurring significant costs. In addition, Table~\ref{table: Existing surveys} provides a comparative study and reachability of existing works that are closely related to the review material presented in this article. The majority of existing research focuses on supply chain sustainability using blockchain and DT as enabling technologies; however, only a limited number of surveys focus primarily on blockchain and DTs as enablers for sustainable BDM. Only a few of them have written in parts about the impact of BDM, while most of the surveys (as seen in Table~\ref{table: Existing surveys}) relate blockchain and DT directly to SSCM. Esmaeilian \textit{et al.}~\cite{ESMAEILIAN2020105064} have covered different business models which are related but distinct to the impacts of technology on the basis of decision-making. For example, their survey introduces social manufacturing - a growing business model that involves collaboration through social media and crowd-sourcing - but the impact is not discussed.

\begin{figure}[ht]
    \centering
    \begin{minipage}{0.6\textwidth}
        \resizebox{9cm}{!}{
            \begin{tikzpicture}
                \begin{axis}[    
                    ylabel=$Article$,
                    xlabel=$Year$,
                    ymin=0,
                    ymax=25,
                    xmin=0,
                    xmax=6,
                    ytick={1,2,3,4,5,6,7,8,9,10,11,12,13,14,15,16,17,18,19,20,21,22,23,24,25},
                    xtick={1,2,3,4,5,6},
                    width=14cm,
                    xticklabels={2020,2021,2022,2023,2024},
                    yticklabels={Kouhizadeh \textit{et al.}~\cite{KOUHIZADEH2021107831},Leng \textit{et al.}~\cite{LENG2020110112},Esmaeilian \textit{et al.}~\cite{ESMAEILIAN2020105064},Kopyto \textit{et al.}~\cite{KOPYTO2020120330},Lee \textit{et al.}~\cite{lee_intelligent_2020}, Guo \textit{et al.}~\cite{GUO9279323},Nguyen \textit{et al.}~\cite{NGUYEN2022108381},Kamble \textit{et al.}~\cite{KAMBLE2022121448},Lohachab \textit{et al.}~\cite{LOHACHAB10.1145/3460287},Sahoo \textit{et al.}~\cite{SAHOO}, Yang \textit{et al.}~\cite{YANG9916158},Sahoo \textit{et al.}~\cite{SAHOO9908297}, Yousefi and Tosarkani~\cite{YOUSEFI10005278}, Rejab \textit{et al.}~\cite{REJEB2023200126}, Yontar \textit{et al.}~\cite{YONTAR2023117173},Singh \textit{et al.}~\cite{SINGH2023109172}, Hammi \textit{et al.}~\cite{HAMMI10.1145/3588999}, Li \textit{et al.}~\cite{LI2023122794}, Adu-Amankwa \textit{et al.}~\cite{ADUAMANKWA2023105064}, Liu \textit{et al.}~\cite{LIU2024248}, Ghasmei \textit{et al.}~\cite{GHASEMI2024100599}, Es-haghi \textit{et al.}~\cite{ESHAGHI2024107342}, Hafeez \textit{et al.}~\cite{HAFEEZ2024107796}, Figueiredo \textit{et al.}~\cite{FIGUEIREDO2024111018}}, 
                    xticklabel style={rotate=90,anchor=east}
                ]
                \addplot[color=black,mark=*] coordinates {
                    (1,1)
                    (1,2)
                    (1,3)
                    (1,4)
                    (1,5)
                    (1,6)
                    (2,7)
                    (2,8)
                    (2,9)
                    (2,10)
                    (3,11)
                    (3,12)
                    (3,13)
                    (3,14)
                    (3,15)
                    (4,16)
                    (4,17)
                    (4,18)
                    (4,19)
                    (5,20)
                    (5,21)
                    (5,22)
                    (5,23)
                    (5,24)
                };
                \end{axis}
            \end{tikzpicture}
        }
    \end{minipage}
    \caption{Timeline of research articles published in similar (overlapping) domains.}
    \label{fig: Published Years}
\end{figure}

Leng \textit{et al.}~\cite{LENG2020110112} have highlighted the importance of the 17 SDGs, thereby conducting a literature review on sustainability across product lifecycle management and sustainable manufacturing as well as identifying the benefits of technological integration. Nguyen~\cite{NGUYEN2022108381} have identified the benefit of DT across multiple applications in a supply chain and proved sustainability by presenting the benefits of DT across different stages of the production process. Reviews in~\cite{KAMBLE2022121448} and \cite{SINGH2023109172} follow a similar structure by presenting the sustainability of the supply chain through various avenues of implementation for a production process. Singh \textit{et al.}~\cite{SINGH2023109172} have discussed the benefits of supply chain resilience. Kopyto \textit{et al.}~\cite{KOPYTO2020120330} approached their empirical study based on reviews from an expert panel of 108 professionals; the results from the panel identify the interoperability issues with blockchain, yet there is strong confidence that by 2035, blockchain will overrule the processes enabling an SSCM. 

Weking \textit{et al.}~\cite{WEKING} have realised the need for adopting business models with blockchain. Therefore, an empirical study was conducted with 99 firms to understand the different business models based on blockchain features that positively impact the firm. It was highlighted by Sahaoo \textit{et al.}~\cite{SAHOO} that much of the existing literature reviews around blockchain and sustainability have not focused on specific sector insights that play a role in adopting blockchain for SSCM. Therefore, Sahoo \textit{et al.} presented a systematic literature review based on research publications using a ranking system to evaluate blockchain effect on SSCM. Adu-Amankwa \textit{et al.}~\cite{ADUAMANKWA2023105064} have focused on building lifecycle management and how sustainability can be achieved using blockchain and DT during pre-and post-construction phases. Lee \textit{et al.}~\cite{lee_intelligent_2020} looked into rising concerns about the maintenance operations of machines within modern manufacturing systems, which impact supply chains. However, with the introduction of industry 4.0 technologies, they have highlighted the use of these technologies to further improve intelligent maintenance operation systems. Li \textit{et al.}~\cite{LI2023122794} have explored the barriers with Industry 4.0 for Industrial Internet of Things devices, particularly with blockchain and DT. They identified 25 barriers, which researchers in this research domain evaluated. Based on the responses and criteria, these are ranked using a decision-making approach highlighting how to overcome the barriers to adoption. Liu \textit{et al.}~\cite{LIU2024248} have evaluated the effectiveness of applying blockchain with federated learning for the application of DT and evaluated it based on criteria such as security, fault-tolerance, fairness, efficiency, cost-savings, profitability and support for heterogeneity. The survey identified that blockchain-based federated learning still has open issues, and a DT enabled by blockchain-based federated learning is still in its infancy. Hafeez \textit{et al.}~\cite{HAFEEZ2024107796} identified the importance of achieving sustainability in industrial furnaces by reducing greenhouse emissions and utilising energy-efficient and renewable energy fuels. DTs have been identified as an enabling technology to foster this transition, and the survey covers the benefits and challenges for organisations to implement DT, which in turn will contribute to achieving sustainability.

Ghasmei \textit{et al.}~\cite{GHASEMI2024100599} have focused on simulation optimisation applied to production scheduling within the application of manufacturing systems. The survey realised several uncertainties within systems, such as random machine breakdowns, order cancellations, and random processing times, which impact scheduling routines. Ghasmei \textit{et al.} highlighted the use of Industry 4.0, particularly the benefit of DT, which allows for real-time reactive scheduling through real-time production monitoring and visualisations. Applications of DTs have been widely studied in literature as they drive sustainability, as highlighted above. However, Es-haghi \textit{et al.}~\cite{ESHAGHI2024107342} highlighted the importance of reducing the computational demand for DTs. To achieve this, Es-haghi \textit{et al.} compared various techniques and methods DT uses for continuous monitoring and evaluation. This includes machine learning techniques such as supervised, unsupervised, reinforcement learning, partial differential equations integration with machine learning and model-ordered reduction methods. Figueiredo \textit{et al.}~\cite{FIGUEIREDO2024111018} examined the life cycle sustainability assessment for building engineering and explored how blockchain and DT can improve sustainability within this industry. For example, sustainability factors like trust and collaboration among stakeholders are strengthened through blockchain, and informed decision-making and maintenance strategies are monitored through DTs. The state-of-the-art solutions focusing on sustainable BDM enabled by blockchain and DT are highlighted in Table~\ref{table: Existing work_summary}.
\begin{table*}[!h]
\tiny
\centering
\resizebox{18cm}{!}{
   % \small
    \begin{tabular}{|p{2cm}|c|p{4cm}|p{5cm}|}
    \hline
    Author (Year) & Ideology & Application & Features \\
    \hline
    [Celik \textit{et al.} 2021]~\cite{CELIK9570246} & Framework & Traceability of models in Building Information Modelling & \vspace{-2.5pt}\begin{itemize}[nosep, leftmargin=*, after={\vspace{-5pt minus 2pt}}] \item Smart contract integration for data interoperability between DT and blockchain \item Access control \end{itemize}\\
    \hline
    [Ding \textit{et al.} 2024]~\cite{DING2024606} & Architecture & A decentralised approach to stakeholder negotiation using smart contracts and with DT & \vspace{-2.5pt}\begin{itemize}[nosep, leftmargin=*, after={\vspace{-5pt minus 2pt}}] \item  Cached DT for online/offline data visualisation \item Two layer verification through smart contracts and IOTA \end{itemize}\\
    \hline
    [Gai \textit{et al.} 2022]~\cite{GAI9833297} & Algorithm & Energy and time optimisation in blockchain-based DT & \vspace{-2.5pt}\begin{itemize}[nosep, leftmargin=*, after={\vspace{-5pt minus 2pt}}] \item Intelligent switch operation between algorithms \item Categorised transaction type \item Strong-weak consensus model \end{itemize} \\
    \hline
    [Ho \textit{et al.} 2021]~\cite{HO2021115101} & Architecture & Aircraft spare part inventory management using blockchain and DT & \vspace{-2.5pt}\begin{itemize}[nosep, leftmargin=*, after={\vspace{-5pt minus 2pt}}] \item Multiple trusted channels for data sharing \item Specific defined data model \end{itemize}\\
    \hline
    [Leng \textit{et al.} 2023]~\cite{LENG10086551} & Framework & Improved resilience using a blockchain smart contract system as a DT & \vspace{-2.5pt}\begin{itemize}[nosep, leftmargin=*, after={\vspace{-5pt minus 2pt}}] \item Bi-level computing architecture is presented for smart contract deployment \item Deep learning model is included within the upper-level smart contracts \end{itemize}\\
    \hline
    [Mandolla \textit{et al.} 2019]~\cite{MANDOLLA2019134} & Framework & Sustainability in additive manufacturing with blockchain and DT & \vspace{-2.5pt}\begin{itemize}[nosep, leftmargin=*, after={\vspace{-5pt minus 2pt}}] \item Smart contract integration \item Proof of data validity \end{itemize}\\
    \hline
    [Monek \textit{et al.} 2024]~\cite{Monek_Fischer_2024} & Architecture & Using DT to minimise error handling and scheduling tasks within a cyber-physical production system & \vspace{-2.5pt}\begin{itemize}[nosep, leftmargin=*, after={\vspace{-5pt minus 2pt}}] \item Fuzzy inference system with the DT decision-making module to map non-linear relationships between inputs and outputs \item Real-time operational adjustments and decision support \end{itemize}\\
    \hline
    [Thanh Le \textit{et al.} 2022]~\cite{THANH10067396} & Framework & Data quality assurance of DT in the supply chain using blockchain & \vspace{-2.5pt}\begin{itemize}[nosep, leftmargin=*, after={\vspace{-5pt minus 2pt}}] \item Model-based simulation approach \item Access control \end{itemize}\\
    \hline
    [Wang \textit{et al.} 2020]~\cite{WANG9384115} & Architecture & Process monitoring and risk management for Industrial Hemp Supply Chain & \vspace{-2.5pt}\begin{itemize}[nosep, leftmargin=*, after={\vspace{-5pt minus 2pt}}] \item Two-layer blockchain - sharding \item Data simulation based on game theory \end{itemize} \\
    \hline
    [Zhang \textit{et al.} 2024]~\cite{ZHANG2024102404} & Framework & Achieved trusted synchronisation of production logistics within a blockchain and digital twin environment, covering the production-distribution-warehouse operations & \vspace{-2.5pt}\begin{itemize}[nosep, leftmargin=*, after={\vspace{-5pt minus 2pt}}] \item Follows the product service system business model to integrate blockchain and DT \item Improve dynamic control and collaborative efficiency in production logistics \end{itemize}\\
    \hline
    \end{tabular}
}
\caption{A summary of existing work on sustainable BDM based on blockchain and/or DT.}
\label{table: Existing work_summary}
\end{table*}

\section{Taxonomy of sustainability in BDM}\label{sec: Section 3}

% \section{Taxonomy of sustainability in BDM and research aspects}
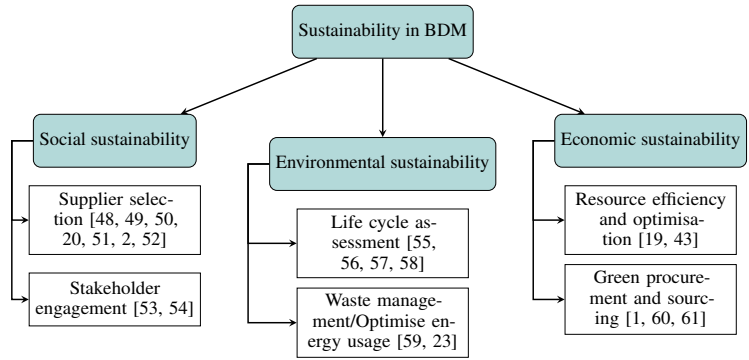
\begin{figure*}
    \centering
    \resizebox{0.6\textwidth}{!}{
    
    \begin{tikzpicture}[node distance=1.5cm]
    
        \tikzstyle{decision} = [rectangle, rounded corners, minimum width=3cm, minimum height=1cm,text centered, draw=black,fill=teal!30]
        \tikzstyle{rect} = [rectangle, minimum width=3cm, minimum height=1cm, text width=3cm, text centered, draw=black]
        \tikzstyle{arrow} = [thick,->,>=stealth]
        
        \node (start) [decision] {Sustainability in BDM};
        \node (child1) [decision, below left of=start, xshift=-4cm, yshift=-1cm] {Social sustainability};
        \node (child2) [decision, below of=start, yshift=-1cm] {Environmental sustainability};
        \node (child3) [decision, below right of=start, xshift=4cm, yshift=-1cm] {Economic sustainability};
        \node (rect1) [rect, below of=child1, yshift=-0cm] {Supplier selection~\cite{RAJESH9940231, ROSTAMNEZHAD2020, ECER2020121981, KELLNER2019505, AHMADI2020123261, KOC2023108820, CHEN.2236, HEZAM, ADITI2024139635}};
        \node (rect2) [rect, below of=child1, yshift=-1.5cm] {Stakeholder engagement~\cite{IBRAHIM8079773, MENDOZA2013, CHEN2024142606}};
        \node (rect4) [rect, below of=child2, yshift=-0cm] {Life cycle assessment~\cite{FIGUEIREDO2021107805, YANG1943-7889.0002028, SANTOS2020121705, ZHANG2020104512, FIGUEIREDO2024111018}};
        \node (rect5) [rect, below of=child2, yshift=-1.5cm] {Waste management/Optimise energy usage~\cite{NIKAS2019140, TSAO2021110452, DIVYAdoi:10.1080/13675567.2021.2014430, KUO, GULIA}};
        \node (rect7) [rect, below of=child3, yshift=-0cm] {Resource efficiency and optimisation~\cite{KIR2021101639, GAI9833297, DISSANAYAKE2018102, LAI6198325}};
        \node (rect8) [rect, below of=child3, yshift=-1.5cm] {Green procurement and sourcing~\cite{GUPTA2017242, LAHRI2021114373, ZHOU2020120950, SHOAIB}};

        \draw [arrow] (start) -- (child1);
        \draw [arrow] (start) -- (child2);
        \draw [arrow] (start) -- (child3);
        \draw [arrow] (child1.west) -- ++(-0.4cm,0) |- (rect1.west);
        \draw [arrow] (child1.west) -- ++(-0.4cm,0) |-  (rect2.west);
        \draw [arrow] (child2.west) -- ++(-0.4cm,0) |- (rect4.west);
        \draw [arrow] (child2.west) -- ++(-0.4cm,0) |-  (rect5.west);
        \draw [arrow] (child3.west) -- ++(-0.4cm,0) |-  (rect7.west);
        \draw [arrow] (child3.west) -- ++(-0.4cm,0) |-  (rect8.west);
    
    \end{tikzpicture}
    }
    \caption{A broad classification of sustainability features in BDM.}
    \label{fig: Classification of sustainability in BDM}
\end{figure*}

This section details the social, environmental and economic sustainability features in BDM that impact responsible computing within organisations and beyond, as illustrated in Fig. \ref{fig: Classification of sustainability in BDM}. \\

\subsection{Social Sustainability}
In BDM, social sustainability refers to an organisation's capacity to conduct socially responsible and equitable operations for all stakeholders. This includes ensuring that the group's actions and choices do not harm society while fostering advancement, inclusivity and diversity. When making decisions in BDM, social sustainability enables stakeholders to take more well-informed and reasonable actions that account for the benefits and social consequences arising from these choices~\cite{KOC2023108820}.

\subsubsection{Supplier selection}
Effective partnerships are vital with suppliers that provide the initial material upstream in the supply chain, and supplier selection has traditionally been seen as one of the most crucial decisions for SSCM~\cite{KOC2023108820}. The solutions here have been based on improving simulation techniques to make informed decisions quickly, where decision models are improved depending on the features. Ahmadi \textit{et al.}~\cite{AHMADI2020123261} have presented a Best-Worst Method (BWM) combined with a modified version of the Preference Ranking Organisation Method for Enrichment of Evaluations (PROMETHEE), which distinguishes between the current state or performance and the desired state. The model is based on a unique parameter called \textit{sustainable innovativeness}. Koc \textit{et al.}~\cite{KOC2023108820} introduced a framework that extends the current Sustainable Supplier Selection (SSS) process; they include innovation, lean principles and knowledge management which utilised existing SSS data to generate further insights for companies to make informed decisions. In addition, a Monte Carlo (MC) assisted, hybrid multi-criteria decision-analysis approach was proposed in their work. It has been highlighted that with existing models, there are chances for a supplier to be selected even if it does not perform well in areas related to social sustainability, as the majority of multi-criteria-based selection methodologies use compensatory aggregation, where good performances in other areas can offset poor performances in one criterion. Rajesh and Aljabhan~\cite{RAJESH9940231} have tackled this compensatory aggregation issue and proposed a new solution using a 2-layered modelling structure to consider social sustainability.
Kellner \textit{et al.}~\cite{KELLNER2019505} have identified the need for considering sustainability factors with the multi-criteria supplier selection problem. Multi-objective optimisation is particularly suitable when approaching the supplier selection problem from the perspective of portfolio configuration. This aims to balance sustainability objectives with the traditional supplier selection goals of cost and supply risk minimisation and service maximisation. The contribution from this research identifies risk and sustainability as two features of BDM~\cite{KELLNER2019505}. Another approach follows the fuzzy best-worst method (F-BWM), capable of better modelling human thought and extracting SSCM practices. This is enhanced by using a conventional Combined Compromise Solution with the Bonferroni mean function (CoCoSo'B) to identify the most appropriate supplier for the supply chain~\cite{ECER2020121981}. Rostamnezhad \textit{et al.}~\cite{ROSTAMNEZHAD2020} have analysed the different aspects affecting social sustainability. Their results were based on expert interviews, and the factors were divided into five sub-models, including stakeholders' engagement, workforce needs consideration, safety-related factors,
health-related factors and management considerations. This data presents a new hybrid system dynamic (SD)-fuzzy Decision-Making Trial and Evaluation Laboratory (DEMATEL) method. Chen \textit{et al.}~\cite{CHEN.2236} have proposed integrating Total Interpretive Structural Modelling (TISM) and the Fuzzy Analytic Network Process (FANP) method to create a score for selecting a socially responsible supplier. The TISM technique can explain the interrelationships among evaluation criteria and present a structured and visual system. The FANP method can calculate the final score to determine the most appropriate supplier. Hezam \textit{et al.}~\cite{HEZAM} have proposed combining Spherical Fuzzy Set (SFS)-Entropy to handle vague, uncertain data and Combinative Distance-Based Assessment (CODAS), which is a reliable Multi-Criteria Decision Model (MCDM), to rank the alternatives. They used a DT to help determine the most suitable supplier. Aditi~\textit{et al.}~\cite{ADITI2024139635} have used industrial experts to identify barriers that influence supplier selection Sustainable Manufacturer-Supplier Collaboration (SMSC), ranked into five levels based on their significance, for example, fear of failure for sustainable collaboration adoption is classified as a least significant barrier, while lack of consistent and adequate performance measurement system and unwillingness to share risk and rewards are critical barriers that prevent SMSC. To classify these barriers, they used Total Interpretive Structural Modelling with DEMATEL.
\subsubsection{Stakeholder Engagement}
Building trust ties with stakeholders makes it possible to address complaints and changes at the outset of a transformation project. By involving them, potential issues with the change process - like ignoring crucial local demands and interests - can be avoided, improving social sustainability~\cite{IBRAHIM8079773}. Ibrahim \textit{et al.}~\cite{IBRAHIM8079773} have focused on smart, sustainable cities and created a model considering the stakeholders' effective and efficient engagement process. Mendoza and Clemen~\cite{MENDOZA2013} have explored stakeholder engagement towards outsourcing sustainable practices between buyers and sellers. Their research followed a game theory approach to incorporate the impact of stakeholder pressure towards process improvement and quality decisions in the supply chain between buyers and sellers. Chen \textit{et al.}~\cite{CHEN2024142606} have used a case study of sustainable farming and realised that there is a limitation in the adoption process, noting that research often overlooks other stakeholders involved and lacks the application of evolutionary game theory to support technology promotion. This insight is essential for developing targeted intervention strategies that address the specific needs of the agricultural community and highlight the need for refined subsidy policies and strategic planning to enhance their effectiveness.
\subsection{Environmental Sustainability}
Due to the considerable resource impact that businesses have, environmental sustainability is vital in BDM. The operations of businesses can hurt the environment by causing pollution, the depletion of natural resources, and greenhouse gas emissions ~\cite{YANG1943-7889.0002028, SANTOS2020121705}. Climate change and biodiversity loss are just two examples of the long-term effects that these environmental problems may have on the globe. Therefore, taking environmental considerations into account in business decision-making can aid organisations in minimising their adverse effects, encouraging resource efficiency and ensuring the long-term viability of their operations. Furthermore, businesses prioritising environmental sustainability may be better positioned to comply with legal obligations, attract eco-aware clients and enhance their standing as socially responsible companies.
\subsubsection{Life Cycle}
By minimising materials, organisations can significantly change the carbon footprint created throughout their life cycle. Figueiredo \textit{et al.}~\cite{FIGUEIREDO2021107805} have discussed the prioritisation of building life-cycle sustainability. It is stated that large-scale structures must maintain safety, reduce failure risks, and introduce a new decision-modelling framework, with a Fuzzy Analytic Hierarchy Process chosen as the framework. Yang~\cite{YANG1943-7889.0002028} identified the issues with existing adaptive risk-based life-cycle management systems and introduced a deep reinforced learning model. The proposed solution helps reduce Life Cycle Costs (LCC), contributing to resource consumption, emissions, waste generation and pollution that, as a result, supports environmental sustainability. Santos \textit{et al.}~\cite{SANTOS2020121705} have identified the relationship between the impact of Life Cycle Assessment (LCA) and LCC on sustainability to maximise the life cycle of a product. 
\subsubsection{Optimise Energy Usage}
Nikas \textit{et al.}~\cite{NIKAS2019140} have identified the ongoing transparency issues within existing models that impact policymakers. Their research followed the Fuzzy Cognitive Mapping (FCM) technique to solve this transparency problem. The simulations from FCM are based on qualitative and quantitative values, including various parameters such as energy consumption, efficiency, the tariff for building integration and economic growth, which help quantify risk-related scenarios. The results show that FCM is a viable soft computing technique to disseminate the most suitable energy-efficient mechanism, thereby improving sustainability in decision modelling. Tsao and Thanh~\cite{TSAO2021110452} have identified energy losses in microgrids and proposed a P2P energy-saving solution to resolve them using a genetic algorithm that adopts blockchain. This supports security and sustainability, as well as allowing participants to take control of the energy system without the need for a central regulatory authority. Choudhary and Kumar \cite{DIVYAdoi:10.1080/13675567.2021.2014430} have highlighted the importance of a circular economy, for example, the end-of-utilisation of a product, which could lead to additional business transactions due to renovation and remanufacturing. They have identified risks with circular economy to facilitate supply chain management with adoption strategies. The study uses Kappa, a statistical analysis tool, and Interpretive Structural Modelling (ISM) to determine the interrelationships among the risks in evaluating the proposed taxonomy of risks. Kuo \textit{et al.}~\cite{KUO} discussed the importance of effective waste management for sustainable growth. However, with effective waste management approaches, there is an additional incurred cost, which could lead to financial problems for organisations. Kuo \textit{et al.} have highlighted that effectively minimising increased cost can be achieved by identifying critical indicators instead of taking action on all indicators. The use of blockchain has been proposed and evaluated using a fuzzy three-step decision-making model based on AI, Technique for Order Preference by Similarity to Ideal Solution (TOPSIS) and Vise Kriterijumska Optimizacija I Kompromisno Resenje (VIKOR) methodology. Gulia \textit{et al.}~\cite{GULIA} have highlighted the importance of environmental sustainability for supply chains. A mathematical model is formulated to achieve three objectives: waste minimisation, profit maximisation and penalty cost minimisation. Fuzzy set theory is used to handle ambiguity, and LINGO software is used to determine the optimal solution.
\subsection{Economic Sustainability}
For several reasons, economic sustainability is significant in BDM. For example, it ensures a business can sustain long-term financial viability and profitability~\cite{YANG1943-7889.0002028}. This is essential for the survival and expansion of an organisation and is advantageous to all of its stakeholders - including staff, clients and shareholders. As a result, companies may contribute to developing a more sustainable economic system that benefits all parties by adopting decisions that promote economic growth and stability.
Economic sustainability is directly related to social and environmental sustainability. For instance, businesses that make investments in environmentally friendly technologies and business practices may be able to lower their operational expenses over time, which benefits both their production line and the environment ~\cite{SHAW2016483}. 

\subsubsection{Resource Efficiency and Optimisation}
With digital transformation, optimisation and efficiency have contributed to BDM through the quality of service and cost efficiency, which affect sustainable practices. Kir and Erdogan~\cite{KIR2021101639} introduced AgileBPM as a complete knowledge-intensive process management solution. The research has identified the need for a more optimised and efficient Business Process Management (BPM) where the ultimate goal of BPM systems shifts from maintaining process automation to assisting business experts in decision-making. The optimisation problem with unknown decision parameters is an issue with the current MCDMs. An MCDM based on the Markov chain is proposed by Fu \textit{et al.}~\cite{FU2022108436}, which achieves sustainability by improving resource efficiency. Simultaneous Evaluation of Criteria and Alternatives (SECA) and double Normalisation-based Multiple Aggregation (DNMA) are examples of newer MCDM techniques that can explored to avoid drawbacks in subjective opinions in the decision-making process enabling sustainability, as suggested in~\cite{ECER2020121981}. Dissanyake and Cross~\cite{DISSANAYAKE2018102} have highlighted the importance of supply chain operations reference, which was defined by the supply chain council. It consists of strategic parameters and objectives to maintain reliability, flexibility, agility, costs and assets to drive supply chain excellence and performance. They identified the importance of supply chain performance and proposed a systematic mechanism to identify factors that impact the supply chain. Lai \textit{et al.}~\cite{LAI6198325} have discussed the impact of Mass Customisation Capability (MCC) on environmental and economic sustainability; in particular, the research focused on the effects of internal integration, customer integration and supplier integration. Although supplier integration did not contribute to the development of the MCC, the former two factors did, particularly when demand is uncertain and completion is intense.
\subsubsection{Green Procurement and Sourcing}
Sustainable requirements can be integrated into the supply chain by using green innovation capabilities as a critical strategy enabled by key facilitators such as knowledge sharing, fairness perception and embeddedness~\cite{ZHOU2020120950}. In addition, Supplier Green Image (SGI) is a key factor that impacts economic green sourcing in a supply chain. However, SGI is difficult to quantify, which can interfere with BDM sustainability and affect suppliers evaluated on SGI~\cite{LAHRI2021114373}. Lahri \textit{et al.}~\cite{LAHRI2021114373} implemented a new decision model that takes into account SGI and Supply Chain Network Design (SCND). The framework is split into two stages where the information is quantified using BWM, TOPSIS, fuzzy possibility theory and the $\mathbf{\epsilon}$-constrained method. The results showed how the decision model could generate cost-effective carbon management strategies. Gupta and Barua have introduced a three-phase methodology~\cite{GUPTA2017242} focusing on Small and Medium Enterprises (SMEs) to identify self-reliant and economically stable suppliers when adjusted to green sourcing. Shoaib \textit{et al.}~\cite{SHOAIB} have used the triple bottom line approach to identify developing green logistics. Fifteen expert participants were involved in the study to identify 21 different enablers. Management support, government regulation and legislation and public and consumer pressure were prioritised. This was first evaluated through Exploratory Factor Analysis (EFA), followed by fuzzy DEMATEL to understand the interrelationships among enablers, and finally, the ISM was used to form the framework.

\subsection{Summary}
Table~\ref{table: Sustainability in BDM} highlights the state-of-the-art approaches applicable to sustainability in BDM. Sustainability in BDM can be improved by using models that can help with criteria evaluations and performance predictions. Integrating risk assessment is another aspect to explore with sustainability, as it can highlight the resource dependency of the systems. Furthermore, setting up the baselines is essential as this helps compare the solutions and identify the most applicable based on their relevant use cases. 

\section{Taxonomy of Sustainability with Blockchain}\label{sec: Section 4}
\begin{figure}[!h]
  \centering
  \includegraphics[height=6cm, width=1\linewidth]{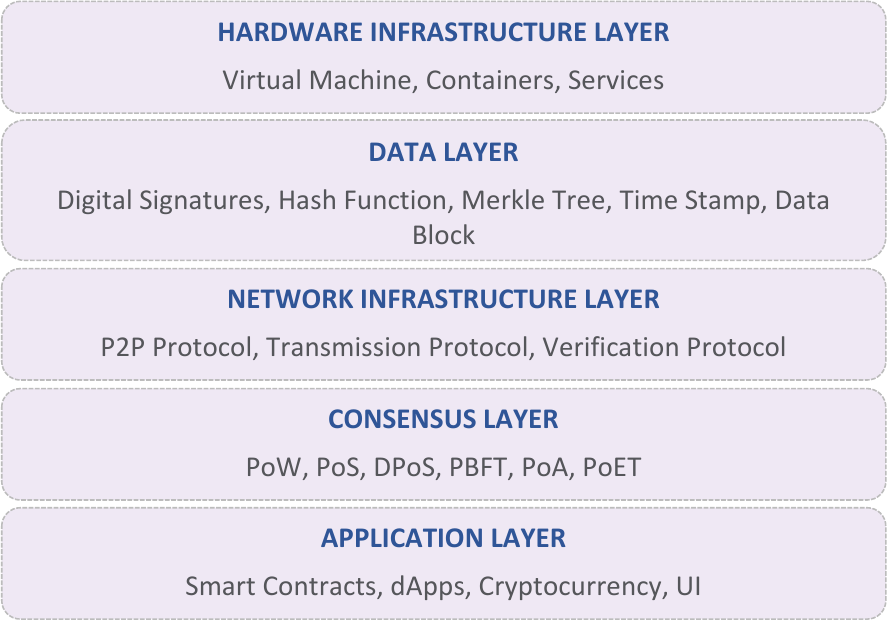}
  \caption{An illustration of the blockchain architecture \cite{LOHACHAB10.1145/3460287}.}
  \label{fig: Blockchain architecture}
\end{figure}

Blockchain is a decentralised distributed ledger that is an append-only data structure and does not depend on a trusted third party to manage its users~\cite{HUANG10.1145/3441692, HERLIHY10.1145/3209623, XIAO8972381}. The different layers of a blockchain are depicted in Fig.~\ref{fig: Blockchain architecture}. A consensus algorithm is executed within the consensus layer, which is the blockchain's premise and acts as a decision-making body to allow fairness across the entire network. It is used within a blockchain architecture (as seen in Fig.~\ref{fig: Blockchain architecture}) to agree on the state of the blockchain between participants~\cite{XIAO8972381, SHARMA8944509}. Interest in blockchain grew with the introduction of the Nakamoto consensus algorithm in 2008, which led to the first crypto-currency known as Bitcoin~\cite{MOHAN10.1145/3299869.3314116, HERLIHY10.1145/3209623, NAKAMOTO2008bitcoin}. Since then, there has been a proliferation of innovation in blockchain, with improvements in consensus protocols impacting performance in throughput, scalability and fault tolerance~\cite{XIAO8972381}. 

\begin{landscape}
    \begin{table}[hb]
        \centering
        \resizebox{23.5cm}{!}{
        \begin{tabular}{|p{2cm}|p{5cm}|c|c|c|p{2cm}|p{2.5cm}|p{1.6cm}|p{2.2cm}|p{2.2cm}|p{2cm}|p{1.8cm}|p{3cm}|}
            \hline
            \multirow{2}{*}{\textbf{Author (Year)}} & \multirow{2}{*}{\textbf{Ideology}} & \multirow{2}{*}{\textbf{Framework models}} & \multicolumn{2}{c|}{\textbf{No. of decision model criteria}} & \multicolumn{7}{c|}{\textbf{BDM properties}} & \multirow{2}{*}{\textbf{Evaluation technique}}\\
            \cline{4-12}
            & & & Criteria & Sub-Criteria & \textbf{Multi-Criteria Evaluation} & \textbf{Performance measurements} & \textbf{Data \& Analytics} & \textbf{Interpersonal uncertainty} & \textbf{Intrapersonal uncertainty} & \textbf{Adaptability} & \textbf{Scalability} &\\
            \hline
            [Aditi \textit{et al.} 2024]~\cite{ADITI2024139635} & Identify barriers for sustainable supplier selection for manufacturing by industrial experts and the importance of these barriers were ranked based on TISM &  \multicolumn{1}{c|}{\begin{tabular}[c]{@{}c@{}} TISM-DEMATEL \end{tabular}} & 19 & {\ding{55}} & \centering {\ding{51}} & \centering {\ding{51}} & \centering {\ding{51}} & \centering {\ding{51}} & \centering {\ding{51}} & \centering {\ding{55}} & \centering {\ding{55}} & Case study - Manufacturing \\
            \hline
            [Ahmadi \textit{et al.} 2020]~\cite{AHMADI2020123261} & Sustainable innovative suppliers within emerging economies & \multicolumn{1}{c|}{\begin{tabular}[c]{@{}c@{}}BWM\\ PROMETHEE \end{tabular}} & 3 & 20 & \centering {\ding{51}} & \centering {\ding{51}} & \centering {\ding{51}} & \centering - & \centering - & \centering {\ding{51}} & \centering - & Case study - Manufacturing\\
            \hline
            [Chen \textit{et al.} 2018]~\cite{CHEN.2236} & Social and environmental index for supplier selection & \multicolumn{1}{c|}{\begin{tabular}[c]{@{}c@{}}TISM\\ FANP \end{tabular}} & 6 & 20 & \centering {\ding{51}} & \centering {\ding{51}} & \centering {\ding{51}} & \centering {\ding{51}} & \centering - & \centering {\ding{55}} & \centering - & Case study - Food industry\\
            \hline
            [Ecer and Pamucar 2020]~\cite{ECER2020121981} & Quantifying vague, imprecise and subjective data & \multicolumn{1}{c|}{\begin{tabular}[c]{@{}c@{}}F-BWM\\ CoCoSo'B \end{tabular}} & 3 & 15 & \centering {\ding{51}} & \centering \centering {\ding{51}} & \centering {\ding{51}} & \centering - & \centering {\ding{51}} & \centering {\ding{55}} & \centering - & Case study - Manufacturing\\
            \hline
            [Figueiredo \textit{et al.} 2021]~\cite{FIGUEIREDO2021107805} & A framework to identify the best building material to maximise the life-cycle of a building & \multicolumn{1}{c|}{\begin{tabular}[c]{@{}c@{}} BIM \\ Fuzzy-AHP \end{tabular}} &  3 & 5 & \centering {\ding{51}} & \centering {\ding{51}} & \centering {\ding{51}} & \centering - & \centering - & \centering {\ding{51}} & \centering {\ding{51}} & Case study - Residential building \\
            \hline
            [Fu \textit{et al.} 2022]~\cite{FU2022108436} & A MCDM that does not require defined parameters &  \multicolumn{1}{c|}{\begin{tabular}[c]{@{}c@{}} Markov Chain \end{tabular}} & - & - & \centering {\ding{51}} & \centering {\ding{51}} & \centering {\ding{51}} & \centering - & \centering - & \centering {\ding{51}} & \centering - & Comparison with TOPSIS and SAW \\
            \hline
            [Gupta and Barua 2017]~\cite{GUPTA2017242} & Supplier selection for SMEs based on green innovation &  \multicolumn{1}{c|}{\begin{tabular}[c]{@{}c@{}} BDM \\ TOPSIS \end{tabular}} & 7 & 42 & \centering {\ding{51}} & \centering {\ding{51}} & \centering {\ding{51}} & \centering - & \centering {\ding{51}} & \centering {\ding{51}} & \centering - & Case study - Automotive \\
            \hline
            [Hezam \textit{et al.} 2024]~\cite{HEZAM} & Overcome uncertainty and incomplete data SFS-Entropy and SFS-CODAS is used for supplier selection, this is also integrated with a DT to suggest the most sustainable supplier & \multicolumn{1}{c|}{\begin{tabular}[c]{@{}c@{}} SFS-Entropy-CODAS \end{tabular}} & 8 & 38 & \centering {\ding{51}} & \centering {\ding{51}} & \centering {\ding{51}} & \centering {\ding{51}} & \centering {\ding{51}} & \centering {\ding{51}} & \centering {\ding{51}} & Case study - Egypt gas firm \\
            \hline
            [Kellner \textit{et al.} 2018]~\cite{KELLNER2019505} & Integrate risk into the supplier selection problem using multi-objective programming & \multicolumn{1}{c|}{\begin{tabular}[c]{@{}c@{}}ANP\\ Markowitz's portfolio model \end{tabular}} & 8 & 22 & \centering - & \centering {\ding{51}} & \centering {\ding{51}} & \centering {\ding{51}} & \centering - & \centering {\ding{51}} & \centering - & Case study - Automotive\\
            \hline
            [Koc \textit{et al.} 2023]~\cite{KOC2023108820} & Implementing Innovation, Lean Principals, and Knowledge Management for sustainable supplier selection & \multicolumn{1}{c|}{\begin{tabular}[c]{@{}c@{}}Monte Carlo\\ AHP \\ TOPSIS \end{tabular}} & 6 & 41 & \centering {\ding{51}} & \centering {\ding{51}} & \centering {\ding{51}} & \centering {\ding{51}} & \centering {\ding{51}} & \centering {\ding{51}} & \centering - & Case study - Construction\\
            \hline
            [Lahri \textit{et al.} 2020]~\cite{LAHRI2021114373} & Incorporate Supplier Green Image in the conventional SCND &  \multicolumn{1}{c|}{\begin{tabular}[c]{@{}c@{}} BDM \\ TOPSIS \\ $\epsilon$- constraint \end{tabular}} & 9 & - & \centering {\ding{51}} & \centering {\ding{51}} & \centering {\ding{51}} & \centering - & \centering {\ding{51}} & \centering {\ding{51}} & \centering - & Case study - Manufacturing \\
            \hline
            [Nikas \textit{et al.} 2018]~\cite{NIKAS2019140} & Bridging transparency between policymakers and systems to improve energy efficiency policies using decision modelling techniques & \multicolumn{1}{c|}{\begin{tabular}[c]{@{}c@{}} FCM \end{tabular}} & - & - & \centering {\ding{51}} & \centering - & \centering {\ding{51}} & \centering - & \centering {\ding{51}} & \centering {\ding{51}} & \centering - & Case study - Greece \\
            \hline
            [Rajesh and Aljabhan 2022]~\cite{RAJESH9940231} & Implementing social sustainability and performance factors to the supplier evaluation process & \multicolumn{1}{c|}{\begin{tabular}[c]{@{}c@{}}Two-Layered GSDM \end{tabular}} & 6 & - & \centering {\ding{51}} & \centering {\ding{51}} & \centering {\ding{51}} & \centering {\ding{51}} & \centering - & \centering {\ding{55}} & \centering - & Case study - Electronics manufacturing \\
            \hline
            [Rostamnezhad \textit{et al.} 2018]~\cite{ROSTAMNEZHAD2020} & A framework based on interactions and factors that affect social sustainability & \multicolumn{1}{c|}{\begin{tabular}[c]{@{}c@{}} SD-fuzzy DEMATEL \end{tabular}} & 5 & 34 & \centering {\ding{51}} & \centering {\ding{51}} & \centering {\ding{51}} & \centering - & \centering {\ding{51}} & \centering {\ding{51}} & \centering - & Case study - Highway project \\
            \hline
            % [Ibrahim \textit{et al.} 2017]~\cite{IBRAHIM8079773} & A model that realises the engagement of stakeholders for smart, sustainable cities & \multicolumn{1}{c|}{\begin{tabular}[c]{@{}c@{}} Stakeholder theory \end{tabular}} & 8 & - & \centering - & \centering - & \centering {\ding{51}} & \centering {\ding{51}} & \centering {\ding{51}} & \centering {\ding{51}} & \centering {\ding{51}} & - \\
            % \hline
            [Santos \textit{et al.} 2020]~\cite{SANTOS2020121705} & A framework that supports automatic analysis of LCA and LCC to support environmental and economic sustainability & \multicolumn{1}{c|}{\begin{tabular}[c]{@{}c@{}} BIM \end{tabular}} & 3 & 5 & \centering - & \centering {\ding{51}} & \centering - & \centering - & \centering - & \centering {\ding{51}} & \centering {\ding{51}} & Case study - Industrial building \\
            \hline
            [Shoaib \textit{et al.} 2022]~\cite{SHOAIB} & A theoretical framework based on 21 enablers which are ranked to identify the most important enablers for green logistics based on the triple bottom line approach &  \multicolumn{1}{c|}{\begin{tabular}[c]{@{}c@{}} EFA \\ fuzzy DEMATEL \\ ISM \end{tabular}} & 21 & {\ding{55}} & \centering {\ding{51}} & \centering {\ding{51}} & \centering {\ding{51}} & \centering {\ding{51}} & \centering {\ding{51}} & \centering - & \centering - & {\ding{55}} \\
            \hline
            [Tsao ad Thanh 2020]~\cite{TSAO2021110452} & Improve energy efficiency in Sustainable Micro-Grids using a P2P fuzzy multi-objective programming model &  \multicolumn{1}{c|}{\begin{tabular}[c]{@{}c@{}} Genetic Algorithm \end{tabular}} & {\ding{55}} & {\ding{55}} & \centering - & \centering - & \centering {\ding{51}} & \centering - & \centering - & \centering - & \centering {\ding{51}} & Case study - Power plant \\
            \hline
        \end{tabular}
        }
        \caption{State-of-the-art approaches applicable to sustainability in BDM ({\ding{51}} Discussed and specified; - Discussed but not specified; {\ding{55}} Not discussed and not specified).}
        \label{table: Sustainability in BDM}
    \end{table}
\end{landscape}

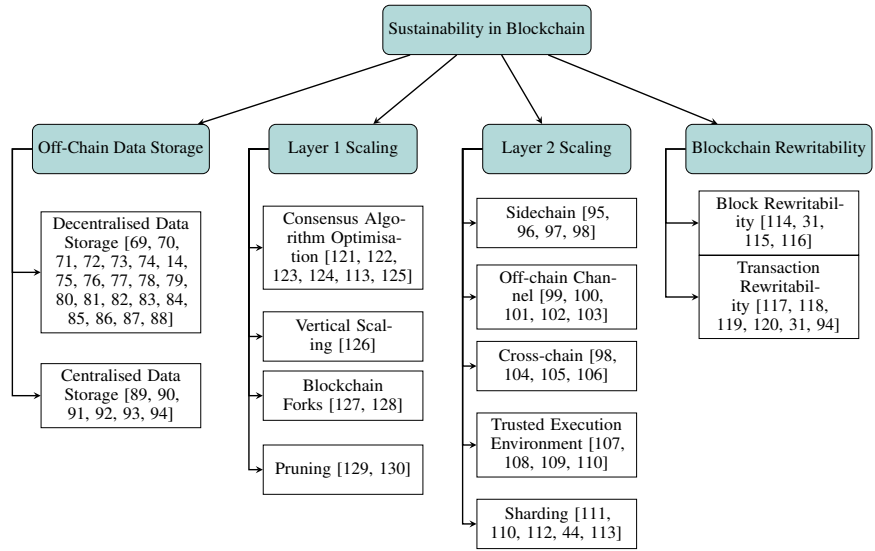
\begin{figure*}
    \centering
    \resizebox{0.7\textwidth}{!}{
        \begin{tikzpicture}[node distance=2cm]
        
            \tikzstyle{decision} = [rectangle, rounded corners, minimum width=3cm, minimum height=1cm,text centered, draw=black,fill=teal!30]
            \tikzstyle{rect} = [rectangle, minimum width=3cm, minimum height=1cm, text width=3cm, text centered, draw=black]
            \tikzstyle{arrow} = [thick,->,>=stealth]
            \node (start) [decision] {Sustainability in Blockchain};
            \node (child1) [decision, below left of=start, xshift=-6cm, yshift=-1cm] {Off-Chain Data Storage};
            % (Off-chain executing)
            \node (child2) [decision, below right of=start, xshift=0cm, yshift=-1cm] {Layer 2 Scaling};
            \node (child3) [decision, below right of=start, xshift=4.5cm, yshift=-1cm] {Blockchain Rewritability};
            % \node (child4) [decision, below right of=start, xshift=3cm, yshift=-1cm] {Access control};
            \node (child5) [decision, below left of=start, xshift=-1.5cm, yshift=-1cm] {Layer 1 Scaling};
            \node (rect1) [rect, below of=child1, yshift=-0.5cm] {Decentralised Data Storage~\cite{KHALID10026822, JAYABALAN2022152, StorJ, RICCI8674173, EMMANUEL8946164, KUMAR9027313, WANG8629877, ZHOU9657488, SAI10009418, YENDURI10.1145/3571306.3571446, KANCHARLA10.1145/3384943.3409435, LI8540048, OZYILMAZ8634961, RAY9319213, HUANG9031420, QIU10.1145/3569966.3569977, SHAHID058674, ARWEAVE, SIA, AUSTRIA9581866, PUTZ2021102425}};
            \node (rect2) [rect, below of=child1, yshift=-3cm] {Centralised Data Storage~\cite{ZHAOLIANG2021124078, ALI8560153, YAMANAKA9838289, GAO10.1145/3581971.3581996, REN2023328, GUO9507330}};
            \node (rect5) [rect, below of=child2, yshift=0.5cm] {Sidechain~\cite{SINGH2020102471, UNNIKRISHNAN10.1007/978-981-19-1018-0_18, ZIEGLER8751308, YIN9543599}};
            \node (rect6) [rect, below of=child2, yshift=-1cm] {Off-chain Channel~\cite{LEE2020101108, CHEN9745518, ZHONG2019327, NEGKA9627997, MAXIMcryptoeprint:2019/352}};
            \node (rect16) [rect, below of=child2, yshift=-2.4cm] {Cross-chain~\cite{YIN9543599, MAO9982450, HE9495937, HEI2022207}};
            \node (rect18) [rect, below of=child2, yshift=-4cm] {Trusted Execution Environment~\cite{MAO10059124, LIND10.1145/3341301.3359627, MILUTINOVIC10.1145/3007788.3007790, DANG10.1145/3299869.3319889, POULAMI235483}};
            \node (rect17) [rect, below of=child2, yshift=-5.6cm] {Sharding~\cite{XU10.14778/3476249.3476283, DANG10.1145/3299869.3319889, YU8954616, WANG9384115, KOKORIS8418625}};
            \node (rect7) [rect, below of=child3, yshift=0.5cm] {Block Rewritability~\cite{ATENIESE7961975, YE10004754, ASHRITHA8728524, DEUBER8835372}};
            \node (rect8) [rect, below of=child3, yshift=-1cm] {Transaction Rewritability~\cite{TIAN10.1145/3427228.3427247, PANWAR10.1145/3450569.3463565, DAVIDcryptoeprint:2019/406, MA9729109, YE10004754, GUO9507330}};
            \node (rect13) [rect, below of=child5, yshift=-0cm] {Consensus Algorithm Optimisation~\cite{CACHIN2017blockchain, KLEINKNECHT9591261, DABBAGH2021102078, FAN9129732, KOKORIS8418625, kokoris-kogias_enhancing_nodate, yin2019hotstuffbftconsensuslens}};
            \node (rect19) [rect, below of=child5, yshift=-1.8cm] {Vertical Scaling~\cite{THAKKAR10.1145/3472883.3486975}};
            \node (rect14) [rect, below of=child5, yshift=-3cm] {Blockchain Forks~\cite{KUBIAK9343117, MADKOUR9477688, NAKAI10444297}};
            \node (rect15) [rect, below of=child5, yshift=-4.5cm] {Pruning~\cite{ABDELHAMID10.1145/3579375.3579420, XU10034838, HUANG10431727}};
            
            \draw [arrow] (start) -- (child1);
            \draw [arrow] (start) -- (child2);
            \draw [arrow] (start) -- (child3);
            \draw [arrow] (start) -- (child5);
            \draw [arrow] (child1.west) -- ++(-0.4cm,0) |- (rect1.west);
            \draw [arrow] (child1.west) -- ++(-0.4cm,0) |-  (rect2.west);
            \draw [arrow] (child2.west) -- ++(-0.4cm,0) |-  (rect5.west);
            \draw [arrow] (child2.west) -- ++(-0.4cm,0) |-  (rect6.west);
            \draw [arrow] (child2.west) -- ++(-0.4cm,0) |-  (rect16.west);
            \draw [arrow] (child2.west) -- ++(-0.4cm,0) |-  (rect18.west);
            \draw [arrow] (child2.west) -- ++(-0.4cm,0) |-  (rect17.west);
            \draw [arrow] (child3.west) -- ++(-0.4cm,0) |-  (rect7.west);
            \draw [arrow] (child3.west) -- ++(-0.4cm,0) |-  (rect8.west);
            \draw [arrow] (child5.west) -- ++(-0.4cm,0) |-  (rect13.west);
            \draw [arrow] (child5.west) -- ++(-0.4cm,0) |-  (rect14.west);
            \draw [arrow] (child5.west) -- ++(-0.4cm,0) |-  (rect15.west);
            \draw [arrow] (child5.west) -- ++(-0.4cm,0) |-  (rect19.west);
        
        \end{tikzpicture}
    }
    \caption{A broad classification of sustainability features in Blockchain.}
    \label{fig: Classification of sustainability in Blockchain}
\end{figure*}

Concerns around sustainability have been raised with the growing applications of blockchain. This includes the electricity consumption for mining blocks and running nodes, scalability, storage and transaction throughput~\cite{XIAO8972381}. Lund \textit{et al.}~\cite{LUND8823906} have highlighted the significance of the 17 SDGs, which correlate with blockchain. For example, Goal 13 addresses~\textit{climate action} and, to tackle this, one of the objectives is to utilise technology extensively to reduce consumption of emissions wherever possible. Munir \textit{et al.}~\cite{MUNIR10.3389/fenrg.2022.899632} have classified the sustainability of blockchain in relation to social, economic and environmental factors. They have identified the benefits of blockchain towards SCM. For example, traceability and increased visibility due to information sharing, openness in procedures and decentralisation are all benefits of blockchain for the economic sustainability of the supply chain. Jan \textit{et al.}\cite{JAN/10.1002/bse.3579} have discussed the role of green supply chain management practices and how blockchain can contribute to support business models to achieve the SDGs. For example, they highlighted that the adoption of blockchain for green warehouses could immediately share transparent information about customer demands, thereby controlling energy utilisation and \(CO_2\) emission, which can help manufacturing industries that emit a lot of greenhouse gases. In addition, Ghobakhloo \textit{et al.}~\cite{GHOBAKHLOO10.1002/bsd2.319} have identified fifteen sustainability functions of blockchain which were mapped to understand the impact of the sustainability functions through which blockchain can contribute to the business environment. For example, Ghobakhloo \textit{et al.} discussed how blockchain drives democracy in supply chains, which promotes sustainable collaboration. To determine the different classifications within a blockchain, Fig.~\ref{fig: Classification of sustainability in Blockchain} categorises literature based on contributions that address sustainability problems in various applications.

\label{sec: Blockchain taxonomy}

\subsection{Off-chain Data Storage}
The need for ample storage was not an initial issue when blockchain was first introduced for cryptocurrency. However, storing large amounts of data is necessary with blockchain's growth across different applications, such as medical, supply chain, IoT and smart grid systems. It is evident that the standard functionality of blockchain for storing large amounts of data affects the performance of the system~\cite{KHALID10026822, DANIEL9684521, LUND8823906}. Due to this, unsustainable resource practices are being used to run the blockchain~\cite{DANIEL9684521}. As a result, there are new implementations to handle the storage of large amounts of data. This paper considers the performance of off-chain storage. With off-chain storage, only the metadata is stored within the blockchain (i.e. recipient and sender details, signatures and hash value), which reduces the load on the blockchain and improves transaction throughput and scalability~\cite{EMMANUEL8946164, DANIEL9684521}. 
\subsubsection{Decentralised Data Storage}
Off-chain decentralised storage promotes sustainability by improving scalability through energy efficiency and reduced infrastructure cost. It encourages community involvement by maintaining data security and resilience through data replication and consistent data accessibility, while avoiding data duplication.
Off-chain decentralised storage promotes sustainability by improving scalability through energy efficiency and reduced infrastructure cost, maintaining data security and resilience by data replication and consistent data accessibility, and avoiding data duplication, which encourages community involvement~\cite{DANIEL9684521, KHALID10026822}. This is a growing area of research, and many blockchain-based applications have considered decentralised storage, such as:
\begin{itemize}
  \item Interplanetary File System (IPFS) - A distributed Peer-to-Peer (P2P) community-driven protocol which integrates concepts from existing popular P2P networks such as BitTorrent, Git and Distributed Hash Tables (DHTs)~\cite{KHALID10026822, BENET2014ipfs}. IPFS is the most popular off-chain data storage method due to its early adoption and robustness~\cite{HENNINGSEN9142766}. Many researchers have used IPFS within applications that contain large amounts of data, with healthcare and supply chain being two key areas~\cite{JAYABALAN2022152, KUMAR9027313, WANG9638510, ZHOU9657488, SAI10009418, YENDURI10.1145/3571306.3571446, KANCHARLA10.1145/3384943.3409435}.
  \item StorJ - A cloud-based platform, albeit with a decentralised P2P architecture that runs through rented space provided by users. It is created following the Satoshi-style blockchain and uses Kademlia DHTs to store and locate data within the network~\cite{StorJ, KHALID10026822, DANIEL9684521}. Research with StorJ is very limited; however, it is a sustainable off-chain storage technique as it provides data redundancy and availability~\cite{YENDURI10.1145/3571306.3571446}. Security evaluations around StorJ have been conducted by the authors of~\cite{RICCI8674173} and~\cite{FIGUEIREDO2021107805} to identify vulnerabilities and extract digital forensic data. 
  \item Swarm - A P2P distributed platform that uses content addressing to store data. Its main objective is to offer decentralised streaming and storage capabilities~\cite{DANIEL9684521, HUANG9031420}. Swarm is an appropriate option if minimal files need to be transferred and exchanged on a decentralised platform. The main benefit of Swarm is that it has incentivisation with low-latency~\cite{RAY9319213}.
   \item Arweave - A sustainable and permanent ledger of knowledge and history which follows a design-based approach. A single upfront fee is required to store files on Arweave, after which the files become part of the consensus~\cite{ARWEAVE}.
  \item Sia - A blockchain-driven platform which facilitates storage contracts, payment transactions and Proof-Of-Storage. Data is distributed across nodes, and peers must generate Proof-Of-Storage periodically to ensure the nodes are valid \cite{SIA, HUANG9031420}. Austria \textit{et al.}~\cite{AUSTRIA9581866} compared Sia with traditional cloud storage regarding cost analysis and performance. The results show that Sia has a lower available time and a high redundancy factor when uploading files; however, Sia had a higher downloading rate. 
\end{itemize}

\subsubsection{Centralised Data Storage}
Centralised data storage to support sustainability within the blockchain is preferred over decentralised storage that has particular applications in relation to storage with large data volumes. It is a traditional storage system that can benefit from performance through reliability and update speed, physical security and improved data preservation~\cite{Dontigney_2019}. Nevertheless, it could be susceptible to certain drawbacks such as maintenance costs and data tampering~\cite{HE10063544}.
\begin{itemize}
  \item Cloud storage - This can increase availability, scalability and management services at low cost by expanding storage resources across several data centres. Specifically, controllability within centralised storage allows access management~\cite{ALI8560153, YAMANAKA9838289}. 
  \item Local database - This allows storing of data within a controlled environment, allowing access control. These are preferred for specific applications. For example, IoT or medical data are stored so they can only be accessed within a local area, which improves real-time data exchange~\cite{YAMANAKA9838289, GAO10.1145/3581971.3581996}.     
\end{itemize}
\subsection{Layer 1 Scaling}
Layer 1 scaling refers to solutions implemented at a blockchain's base layer. It entails changing the underlying blockchain architecture to boost its capacity to process and validate transactions, improve scalability and improve network performance. 
\subsubsection{Consensus Algorithm Optimisation}
Following the research in~\cite{CACHIN2017blockchain, KLEINKNECHT9591261, DABBAGH2021102078}, consensus protocols directly impact energy efficiency and computational overheads that, as a result, affect sustainability. It is important to note that each consensus protocol works differently; therefore, they may not have the same impact on sustainability (see Section~\ref{sec: Section 7} for a summarised comparison of private blockchains). Since consensus protocols have to maintain characteristics of the ledger, which is a challenge, many approaches such as Proof of Work, Proof of Stake and Proof of Authority exhibit different security and performance trade-offs \cite{KOKORIS8418625, KLEINKNECHT9591261, yin2019hotstuffbftconsensuslens}. Kokoris-Kogias \textit{et al.}~\cite{KOKORIS8418625} have implemented a new consensus protocol and blockchain named OmniLedger. OmniLedger is built on ByzCoin~\cite{kokoris-kogias_enhancing_nodate} $-$ their ledger uses sharding to help increase its throughput to match VISA. Unlike other sharding protocols, OmniLedger has achieved cross-shared transaction atomicity, which either succeeds in its entirety or rolls back without any partial changes. 
\subsubsection{Vertical Scaling}
A vertical scaling solution aims to increase the throughput per node, which is achieved by upgrading the hardware of the ledger. Thakkar and Natarajan~\cite{THAKKAR10.1145/3472883.3486975} tested the benefit of vertical scaling. They used a novel scaled-up consensus, Smart Fabric, against Vanilla Fabric and FastFabric - both scaled-up versions of Hyperledger Fabric. Based on the results, a vertical scaling solution does not have a considerable impact. This is because the nature of blockchain is homogeneous $-$ the larger the network scale, the more bandwidth is needed to achieve the desired throughput, which is an indefinite resource~\cite{YU8954616}. 
\subsubsection{Blockchain Forks}
A blockchain that splits into two or more separate chains with a shared history is termed a fork~\cite{MADKOUR9477688}. A blockchain fork can affect sustainability due to resource fragmentation, creating uncertainty among users and increasing energy consumption. Madkour \textit{et al.}~\cite{MADKOUR9477688} approached this issue by implementing an algorithm that splits the block construction into two phases. During the initial phase, the miner mines the block, subsequently adding it to the memory pool; the second phase requires the memory pool to append the block to the chain. Nakai and Shudo~\cite{NAKAI10444297} have discussed how blockchain forks can impact security. This is because to tackle the stability problem with blockchain, batch size of a block should be increased and/or block generation time should be reduced. Therefore, Nakai and Shudo evaluate the two approaches to mathematically prove that block generation time greatly affects fork rate in a proof of work consensus protocol.
\subsubsection{Pruning}
This technique selectively removes data from the blockchain to help improve performance efficiency~\cite{ABDELHAMID10.1145/3579375.3579420}. Abdelhamid \textit{et al.}~\cite{ABDELHAMID10.1145/3579375.3579420} specified three types of pruning techniques: (1) Simple block pruning $-$ where only UTXO remains in the blocks; (2) State-based synchronisation $-$ where only a snapshot of the blockchain UTXO which is temporary is used; and (3) Balance-based synchronisation, where none of the UTXO transactions are tracked before deleting and only the existing accounts and their balances are left. Xu \textit{et al.}~\cite{XU10034838} have defined a more efficient data structure named MRK-Tree that allows for efficient pruning of blocks and experiments to deduce the solution's viability. Huang \textit{et al.}~\cite{HUANG10431727} have realised a large amount of time is being used to synchronise full nodes on the blockchain to improve bootstrapping time further while maintaining the security of data, Huang \textit{et al.} have used a dynamic pruning technique based on features extracted from the blockchain metadata. 
\subsection{Layer 2 Scaling}
Layer 2 solution is synonymously referred to as off-chain execution. It supports sustainability by allowing the primary chain only to process pertinent transactions and ensure that other transactions that could be completed outside are not included~\cite{UNNIKRISHNAN10.1007/978-981-19-1018-0_18}. 
\subsubsection{Trusted Execution Environment (TEE)}
TEE is an operating environment outside the primary system. It provides isolation and protection for running trusted applications. A TEE co-exists with RichOS, which manages system services and resources. TEE is only responsible for code execution - it cannot be influenced by the code's external environment, and ensures security and costs~\cite{MAO10059124}. Different types of TEE can achieve this. For example, Liu \textit{et al.}~\cite{LIU9705115} have used the Arm Cortex-M series TEE microcontroller. As a result, TEE sustainability is achieved with enhanced security and efficient resource utilisation. However, based on the TEE architecture, security is not always adhered to. Poulami \textit{et al.}~\cite{POULAMI235483} have proposed smart contracts executed on the Bitcoin network. They used a TEE to integrate with the network, maintain data security and reduce transaction complexity. Poulami \textit{et al.} implemented the smart contracts with multi-round contracts handling coins, which support real-time application scenarios.
\subsubsection{Sidechain}
Sidechains run in parallel alongside the leading blockchain network. They have their own consensus protocol and rules. Most sidechain frameworks are connected by a two-way peg that allows the bidirectional transfer of assets between the mainchain and sidechain~\cite{SINGH2020102471, UNNIKRISHNAN10.1007/978-981-19-1018-0_18}. Sidechains are classified into two types: parent-child connection and equal connection. In the parent-child connection, one blockchain (called the sidechain) is a `child' of another blockchain (called the mainchain). The sidechain is dependent on the mainchain (at least during the initialisation stage). Two blockchains are treated equally in the equal connection, and either of the two blockchains might be considered the sidechain of the other~\cite{YIN9543599}. Sidechains are a simple technique to improve the scalability of blockchains. On the other hand, sidechains investigate a new method of transitioning blockchain systems without requiring a fork~\cite{YIN9543599}. Singh \textit{et al.} \cite{SINGH2020102471} have presented a review on state-of-the-art side-chain technologies that can support blockchain sustainability.   
\subsubsection{Cross-Chain}
In the context of blockchain, cross-chain refers to the ability to transfer or exchange data between one distributed ledger to another. This can be in the form of assets such as tokens or cryptocurrency. Mao \textit{et al.}~\cite{MAO9982450} defined three ways of establishing cross-chain communication: hash locking, third-party collaboration or using relays/sidechains. Mao \textit{et al.} further explained the communication techniques, conducted an elaborate survey on cross-chain communication and evaluated the use of cross-chain for sustainability blockchain. For example, cross-chains allow interaction and collaboration between blockchain platforms, providing value and information interaction between industry chains. A popular cross-chain system is Polkadot.\footnote{Polkadot: \href{https://polkadot.network}{https://polkadot.network}} 

\subsubsection{Off-chain Channel}
Off-chain channel encompasses two main techniques that solve sustainability issues within a blockchain: 
\begin{itemize}
    \item State channel - A state channel only applies to Turing complete blockchains. For example, when a user interacts with a deployed smart contract on the blockchain, the communication causes a state change, which invokes the user to submit a new transaction. Based on the example, processing multiple transactions each time a smart contract is invoked could affect the performance of the blockchain with respect to time delays, as a block has to be appended for the user~\cite{NEGKA9627997}. The authors in~\cite{NEGKA9627997, MAXIMcryptoeprint:2019/352} elaborated on state channels and their sustainability benefits towards blockchain through taxonomies. 

\onecolumn
\begin{landscape}
        \tiny
        \scriptsize
        \begin{longtable}{|p{2cm}|p{3.5cm}|p{0.8cm}|p{0.6cm}|p{1.8cm}|p{1cm}|p{1.3cm}|p{1.5cm}|p{1.1cm}|p{1cm}|}
            \caption{State-of-the-art approaches applicable for sustainability in the blockchain ({\ding{51}} Discussed and specified; - Discussed but not specified; {\ding{55}} Not discussed and not specified).} \label{table: Sustainability in Blockchain} \\
            \hline
            \multirow{2}{*}{\textbf{Author (Year)}} & \multirow{2}{*}{\centering\textbf{Ideology}} & \multicolumn{5}{c|}{\textbf{Sustainability features in blockchain}} & \multirow{2}{*}{\textbf{Use case}} & \multirow{2}{*}{\textbf{Comparison}} & \multirow{2}{*}{\textbf{Parameters}} \\
            \cline{3-7}
            & & \textbf{Scalability} & \textbf{Security} & \textbf{Consensus mechanism} & \textbf{Adaptability} & \textbf{Computational efficiency} & & &\\
            \hline
            \endfirsthead
            
            \multicolumn{10}{c}
            {{\small\tablename\ \thetable{} -- continued from previous page}} \\
            \hline
            \multirow{2}{*}{\textbf{Author (Year)}} & \multirow{2}{*}{\centering\textbf{Ideology}} & \multicolumn{5}{c|}{\textbf{Sustainability features in blockchain}} & \multirow{2}{*}{\textbf{Use case}} & \multirow{2}{*}{\textbf{Comparison}} & \multirow{2}{*}{\textbf{Parameters}} \\
            \cline{3-7}
            & & \textbf{Scalability} & \textbf{Security} & \textbf{Consensus mechanism} & \textbf{Adaptability} & \textbf{Computational efficiency} & & &\\
            \hline
            \endhead

            [Ali \textit{et al.} 2018]~\cite{ALI8560153} & Securely manage large quantity of data from IoT gateways  & \centering {\ding{55}} & \centering {\ding{51}} & Hyperledger Fabric & \centering - & \centering {\ding{51}} & IoT & \centering - & \multicolumn{1}{c|}{\begin{tabular}[c]{@{}c@{}} Transactions per second \end{tabular}} \\
            \hline
            [Chen \textit{et al.} 2022]~\cite{CHEN9745518} & Extend two-part payment channel to a multi-party payment channel & \centering {\ding{51}} & \centering {\ding{51}} & Ethereum & \centering - & \centering {\ding{51}} & \centering - & \centering {\ding{51}} & \multicolumn{1}{c|}{\begin{tabular}[c]{@{}c@{}} Time for transactions \\ Success ratio \\ Average gas costs \end{tabular}} \\
            \hline
            [Dang \textit{et al.} 2019]~\cite{DANG10.1145/3299869.3319889} & Improve performance in permissioned blockchain settings through sharding and a TEE & \centering {\ding{51}} & \centering {\ding{51}} & Practical Byzantine Fault Tolerance  and Proof of Elapsed Time & \centering {\ding{51}} & \centering {\ding{51}} & \centering - & \centering {\ding{51}} & \multicolumn{1}{c|}{\begin{tabular}[c]{@{}c@{}} Transactions per second \\No. of nodes \\No. of view changes \\ Stale blocks \end{tabular}} \\
            \hline
            [He \textit{et al.} 2022]~\cite{HE9495937} & Smart contract integration for security between cross-chains & \centering - & \centering {\ding{51}} & \centering - & \centering {\ding{51}} & \centering {\ding{51}} & Electric vehicle charging pills & \centering - & \multicolumn{1}{c|}{\begin{tabular}[c]{@{}c@{}} Gas consumption \\ Reputation \end{tabular}} \\
            \hline
            [Hei \textit{et al.} 2021]~\cite{HEI2022207} & Smart contract integration for security between cross-chains & \centering {\ding{51}} & \centering {\ding{51}} & Ethereum & \centering {\ding{51}} & \centering {\ding{51}} & \centering - & \centering - & \multicolumn{1}{c|}{\begin{tabular}[c]{@{}c@{}} Gas cost \\ Fee cost  \end{tabular}} \\
            \hline
            [Huang \textit{et al.} 2024]~\cite{HUANG10431727} & Bootstrap service for a UTXO model of the Bitcoin system based on a pruning strategy with fast synchronisation & \centering {\ding{51}} & \centering {\ding{51}} & Proof of Work & \centering - & \centering {\ding{51}} & \centering Bitcoin Network & \centering {\ding{51}} & \multicolumn{1}{c|}{\begin{tabular}[c]{@{}c@{}} Disk Size \\ Synchronisation Time \end{tabular}} \\
            \hline
            [Jayabalan and Jeyanthi 2022]~\cite{JAYABALAN2022152} & Allow healthcare transfer between institutions without security compromise & \centering {\ding{51}} & \centering {\ding{51}} & Proof of Work & \centering {\ding{51}} & \centering - & Healthcare & \centering {\ding{51}} & \multicolumn{1}{c|}{\begin{tabular}[c]{@{}c@{}} No. of transactions\\ No. of blocks\\ Size\\ Retrieval time \end{tabular}} \\
            \hline
            [Kancharla \textit{et al.} 2020]~\cite{KANCHARLA10.1145/3384943.3409435} & A dependability model based on write and read rates to determine if data is stored off-chain or on-chain & \centering {\ding{51}} & \centering - & \centering - & \centering {\ding{51}} & \centering - & \centering - & \centering {\ding{51}} & \multicolumn{1}{c|}{\begin{tabular}[c]{@{}c@{}} Dependability\\Transaction state \end{tabular}} \\
            \hline
            [Kokoris-Kogias \textit{et al.} 2018]~\cite{KOKORIS8418625} & Handle transactions at the same rate VISA works at by using shards & \centering {\ding{51}} & \centering {\ding{51}} & ByzCoin & \centering {\ding{51}} & \centering {\ding{51}} & \centering - & \centering {\ding{51}} & \multicolumn{1}{c|}{\begin{tabular}[c]{@{}c@{}} Transaction per second \\ Client end-to-end latency \\ Latency \\ Consumed bandwidth  \end{tabular}} \\
            \hline
            [Kumar \textit{et al.} 2020]~\cite{KUMAR9027313} & Handle misuse of patient data and preserve patient privacy & \centering {\ding{51}} & \centering {\ding{51}} & Proof of Work & \centering - & \centering {\ding{51}} & Healthcare & \centering {\ding{51}} & \multicolumn{1}{c|}{\begin{tabular}[c]{@{}c@{}} File size\\ Nodes \end{tabular}} \\
            \hline
            [Li \textit{et al.} 2019]~\cite{LI8540048} & Secure crowdsourcing & \centering {\ding{51}} & \centering {\ding{51}} & Ethereum & \centering {\ding{51}} & \centering {\ding{51}} & Crowdsourcing & \centering {\ding{51}} & \multicolumn{1}{c|}{\begin{tabular}[c]{@{}c@{}} Growth rate of cost \\Transaction confirmation time \end{tabular}} \\
            \hline
            [Li \textit{et al.} 2022]~\cite{LI10026494} & Handle imbalance transaction load to improve sharding throughput & \centering {\ding{51}} & \centering {\ding{51}} & Proof of Work & \centering {\ding{51}} & \centering {\ding{51}} & \centering - & \centering {\ding{51}} & \multicolumn{1}{c|}{\begin{tabular}[c]{@{}c@{}} Transactions per second \\ Average transaction confirmation delay \\ Cumulative distributed function \end{tabular}} \\
            \hline
            [Lind \textit{et al.} 2019]~\cite{LIND10.1145/3341301.3359627} & Work asynchronously and handle malicious nodes through a trusted payment network based on TEE & \centering {\ding{51}} & \centering {\ding{51}} & Proof of Work & \centering - & \centering {\ding{51}} & \centering - & \centering {\ding{51}} & \multicolumn{1}{c|}{\begin{tabular}[c]{@{}c@{}} Throughput \\ Latency \end{tabular}} \\
            \hline
            [Liu \textit{et al.} 2022]~\cite{LIU9705115} & Using a TEE to ensure the security of data transfer between the physical world and blockchain (physical traceability) &  & \centering {\ding{51}} & Tendermint-BFT & \centering {\ding{51}} & \centering {\ding{51}} & Vaccine tracing system & \centering - & \multicolumn{1}{c|}{\begin{tabular}[c]{@{}c@{}} System processing time \\ Transmission time \\ Synchronisation time \end{tabular}} \\
            \hline
            [Moa \textit{et al.} 2023]~\cite{MAO10059124} & Cross-domain authentication between devices located in different IoT domains & \centering {\ding{51}} & \centering {\ding{51}} & \centering - & \centering - & \centering {\ding{51}} & IoT & \centering {\ding{51}} & \multicolumn{1}{c|}{\begin{tabular}[c]{@{}c@{}} Time cost \\ Gas \end{tabular}} \\
            \hline
            [Milutinovic \textit{et al.} 2016]~\cite{MILUTINOVIC10.1145/3007788.3007790} & Implement a consensus protocol with a TEE that is energy and network communication efficient which does not require a synchronisation clock & \centering {\ding{51}} & \centering {\ding{51}} & Proof of Luck & \centering - & \centering {\ding{51}} & \centering - & \centering - & \multicolumn{1}{c|}{\begin{tabular}[c]{@{}c@{}} - \end{tabular}} \\
            \hline
            [Nakai and Shudo 2024]~\cite{NAKAI10444297} & The impact of fork rate based on either increasing block size or reducing block generation time & \centering {\ding{51}} & \centering {\ding{51}} & Proof of Work & \centering - & \centering - & \centering - & \centering - & \multicolumn{1}{c|}{\begin{tabular}[c]{@{}c@{}} Theoretical Fork Rate \\ Block Generation Time \end{tabular}} \\
            \hline
            [Ozyilmaz and Yurdakul 2019]~\cite{OZYILMAZ8634961} & Standardise communication between IoT gateways & \centering {\ding{51}} & \centering {\ding{51}} & Ethereum & \centering {\ding{51}} & \centering {\ding{51}} & IoT & \centering - & \multicolumn{1}{c|}{\begin{tabular}[c]{@{}c@{}} - \end{tabular}} \\
            \hline
            [Qiu and Tian 2022] \cite{QIU10.1145/3569966.3569977} & Trace long-life-cycle of cable manufacturing data & \centering - & \centering {\ding{51}} & Raft with Hyperledger Fabric & \centering - & \centering {\ding{51}} & Supply Chain & \centering {\ding{51}} & \multicolumn{1}{c|}{\begin{tabular}[c]{@{}c@{}} Transactions per second \\ No. of transactions \\ No. of uploads \\ Queries \end{tabular}} \\
            \hline
            [Ray \textit{et al.} 2021]~\cite{RAY9319213} & Produce a privacy scheme between IoT devices sharing data over a network & \centering {\ding{55}} & \centering {\ding{51}} & \centering {\ding{55}} & \centering - & \centering {\ding{51}} & Healthcare & \centering {\ding{51}} & \multicolumn{1}{c|}{\begin{tabular}[c]{@{}c@{}} Upload time \\ Download time \\ Swarm exchange time during upload \\ Swarm exchange time during download \\
            \end{tabular}} \\
            \hline
            [Ren \textit{et al.} 2020]~\cite{REN2023328} & Securely manage large volumes of data with off-chain storage by using a subvector commitment scheme & \centering {\ding{55}} & \centering {\ding{51}} & \centering - & \centering - & \centering {\ding{51}} & Spatio-temporal Data & \centering {\ding{51}} & \multicolumn{1}{c|}{\begin{tabular}[c]{@{}c@{}} Time cost \end{tabular}} \\
            \hline
            [Sai \textit{et al.} 2022]~\cite{SAI10009418} & Handle transactions within a large data set & \centering - & \centering {\ding{51}} & Ethereum & \centering - & \centering - & Food Supply Chain & - & \multicolumn{1}{c|}{\begin{tabular}[c]{@{}c@{}} - \end{tabular}}  \\
            \hline
            [Shadhid \textit{et al.} 2020]~\cite{SHAHID058674} & Ensure traceability and product delivery based on the reputation score of supplier & \centering {\ding{51}} & \centering {\ding{51}} & Ethereum & \centering - &  & Food Supply Chain & \centering {\ding{51}} & \multicolumn{1}{c|}{\begin{tabular}[c]{@{}c@{}} Gas consumption \\ Mining time \end{tabular}} \\
            \hline
            [Thakkar and Natarajan 2021]~\cite{THAKKAR10.1145/3472883.3486975} & Improve scalability of Execute-Order-Validate blockchain algorithms & \centering {\ding{51}} & \centering - & Hyperledger Fabric & \centering - & \centering {\ding{51}} & \centering - & \centering {\ding{51}} & \multicolumn{1}{c|}{\begin{tabular}[c]{@{}c@{}} Throughput \\ Block Processing \end{tabular}} \\
            \hline
            [Wang \textit{et al.} 2020]~\cite{WANG9384115} & A two-layered blockchain architecture with sharding to maintain blockchain performance & \centering {\ding{51}} & \centering {\ding{51}} & Proof of Authority & \centering - & \centering {\ding{51}} & Hemp supply chain risk management & \centering - & \multicolumn{1}{c|}{\begin{tabular}[c]{@{}c@{}} - \end{tabular}} \\
            \hline
            [Wang \textit{et al.} 2022]~\cite{WANG9638510} & Manage personal health records in Internet Medical of Things devices using a consortium blockchain & \centering {\ding{51}} & \centering {\ding{51}} & PBFT with Hyperledger Fabric & \centering {\ding{51}} & \centering {\ding{51}} & Healthcare & \centering {\ding{51}} & \multicolumn{1}{c|}{\begin{tabular}[c]{@{}c@{}} Security properties\\ Communication overheads\\ Computational efficiency  \end{tabular}} \\ 
            \hline
            [Yamanaka \textit{et al.} 2022]~\cite{YAMANAKA9838289} & Implementing a user-centric in-network caching mechanism for off-chain storage & \centering {\ding{51}} & \centering {\ding{51}} & \centering - & \centering - & \centering {\ding{51}} & \centering {\ding{55}} & \centering {\ding{55}} & \multicolumn{1}{c|}{\begin{tabular}[c]{@{}c@{}} Cumulative distributed function \\ Total network traffic \end{tabular}} \\
            \hline
            [Yenduri \textit{et al.} 2023]~\cite{YENDURI10.1145/3571306.3571446} & Improving management of software maintainability practices & \centering - & \centering {\ding{51}} & \centering - & \centering - & \centering - & Healthcare & \centering - & \multicolumn{1}{c|}{\begin{tabular}[c]{@{}c@{}} - \end{tabular}} \\
            \hline
            [Yin \textit{et al.} 2022]~\cite{YIN9543599} & A novel cross-chain certificate generation for data transfer between the sidechain and mainchain & \centering {\ding{51}} & \centering {\ding{51}} & Proof of Stake and Proof of Work & \centering {\ding{51}} & \centering {\ding{51}} & \centering - & \centering {\ding{51}} & \multicolumn{1}{c|}{\begin{tabular}[c]{@{}c@{}} Average time for transaction \\ Certificate size \end{tabular}} \\
            \hline
            [Zhaoliang \textit{et al.} 2020]~\cite{ZHAOLIANG2021124078} & Ensure secure storage of agriculture crop data & \centering {\ding{55}} & \centering {\ding{51}} & \centering {\ding{55}} & \centering - & \centering - & Agriculture & \centering {\ding{51}} & \multicolumn{1}{c|}{\begin{tabular}[c]{@{}c@{}} Storage rate \\ Time \end{tabular}} \\
            \hline
            [Zhong \textit{et al.} 2018]~\cite{ZHONG2019327} & A novel versatile light payment technique to improve performance of blockchain systems by handling high-frequency and high-capacity transactions & \centering {\ding{51}} & \centering {\ding{51}} & \centering {\ding{55}} & \centering {\ding{51}} & \centering {\ding{51}} & \centering - & \centering - & \multicolumn{1}{c|}{\begin{tabular}[c]{@{}c@{}} - \end{tabular}} \\
            \hline
            [Zhou \textit{et al.} 2021]~\cite{ZHOU9657488} & Implement sustainable reward mechanism to influence new users to join and existing users to contribute & \centering - & \centering {\ding{51}} & Ethereum & \centering - & \centering {\ding{51}} & Product Evaluation Management & \centering {\ding{51}} & \multicolumn{1}{c|}{\begin{tabular}[c]{@{}c@{}} Gas\\ Execution time\\ Rewards \end{tabular}} \\
            \hline
        \end{longtable}
\end{landscape}
\twocolumn
 
    \item Payment channel - This is specifically designed to handle off-chain payments. It does not require the blockchain to be Turing complete. Instead, it relies on cryptography techniques and time locks. A current issue in blockchain is the bottleneck caused by payment transactions between two entities. This leads to performance inefficiency and increased costs, affecting the use of the distributed ledger. This can be handled by payment channels~\cite{ZHONG2019327, CHEN9745518}. The most popular payment channels are the Lightning Network\footnote{Lightning network: \href{https://lightning.network}{https://lightning.network}} for Bitcoin and Radien Network\footnote{Radien network: \href{https://raiden.network}{https://raiden.network}} for Ethereum \cite{UNNIKRISHNAN10.1007/978-981-19-1018-0_18}. Lee and Kim \cite{LEE2020101108} tested the robustness of the Lightning network and identified its characteristics and vulnerabilities, highlighting key concerns for the network.  
\end{itemize}
\subsubsection{Sharding} 
When the number of transactions in existing blockchain systems increases, scalability becomes a significant barrier to implementation, which can be handled with sharding. Sharding is also known as horizontal scaling. This technique divides blocks into multiple shards, allowing participating nodes to process and store transactions from a few shards instead of the full node~\cite{YU8954616}. With sharding, the compromising of attackers is limited, which must be handled by assigning nodes to shards securely. Yu \textit{et al.}~\cite{YU8954616} covered an in-depth review of the state-of-the-art sharding mechanisms for blockchain. Dang \textit{et al.}~\cite{DANG10.1145/3299869.3319889} achieved a secure sharding mechanism using a TEE. A sharding protocol must ensure both atomicity and isolation of transactions. The authors identified that although Byzantine Fault Tolerant (BFT) consensus protocols assume a stronger failure model against malicious attackers, they have a scalability bottleneck which affects sustainability. Their research focused on sharding in a permissioned setting based on epochs. Li \textit{et al.}~\cite{LI10026494} identified two main limitations in current sharding techniques: (1) there is still a throughput gap between the existing sharding protocols and the potential throughput, a 30\% drop has been noticed; and (2) there is still a delay from when a user initiates a transaction until they receive the confirmation. The authors pointed out the reasoning for this limitation: cross-sharding and an imbalance transaction load. 
\subsection{Blockchain Rewritability/Redaction}
The concept of blockchain rewritability is still in its early stages, and the immutability feature has posed challenges in certain instances, especially in understanding the reasoning behind such a requirement. 

Panwar \textit{et al.}~\cite{PANWAR10.1145/3450569.3463565} identified the importance of privacy regulations such as the GDPR (General Data Protection Regulation), which directly has an impact on blockchain as sensitive user data is stored in the ledger. Rewritability can also benefit blockchain processes like smart contracts. For example, if an existing smart contract contains flaws, it could be exploited by attacks like the Decentralised Autonomous Organisation ~\cite{ZHAO8248566}, which took 3.6 million ether. This has an impact on the blockchain's long-term sustainability~\cite{YE10004754}. Instead, the smart contract can be rewritten under access control to prevent harmful cyber attacks. The sustainability impact of rewritability covers error correction, regulatory compliance and energy efficiency. However, it is vital to maintain trust in the ledger by following tight access control, authorisation and cryptographic techniques \cite{GUO9507330}. The process of rewritability can follow a cryptographic-based scheme or a non-cryptographic-based scheme. While cryptographic techniques are more computationally intensive, they use digital signatures and hash functions to rewrite data. A non-cryptographic technique would be easier to implement as security is not the primary concern. However, this could compromise the integrity of the blockchain.
RSA-based redaction and voting-based redaction are examples of cryptographic and non-cryptographic techniques, respectively~\cite{YE10004754}. Ye \textit{et al.}~\cite{YE10004754} have discussed the two techniques in-depth and the literature around them.

\subsubsection{Block Rewritability}
Block rewritability replaces an entire block in the distributed ledger. This means all transactions within that block are replaced. Atenises \textit{et al.}~\cite{ATENIESE7961975} initially proposed a Chameleon Hash Function (CHF), which allows only modifying the entire block. Ashritha \textit{et al.}~\cite{ASHRITHA8728524} have used CHF for block predictability and proposed sharing the trapdoor key, which is a secret key that allows redacting a block. Deuber \textit{et al.}~\cite{DEUBER8835372} proposed a consensus-based voting mechanism for redacting a block in Bitcoin.      
\subsubsection{Transaction Rewritability}
With transaction-level rewritability, there is more granularity in adjusting specific content within a block. Yet, with certain encryption techniques, computational power is high. Guo \textit{et al.}~\cite{GUO9507330} have handled this by offloading heavy computation to the cloud. Tian \textit{et al.}~\cite{TIAN10.1145/3427228.3427247} have proposed a framework using a Policy-based Chameleon Hash Function with a black box for the trapdoor key. Panwar \textit{et al.}~\cite{PANWAR10.1145/3450569.3463565} have presented a revocable chameleon hash with an ephemeral trapdoor scheme with Attribute-Based Encryption.

\subsection{Summary}
Sustainability within the blockchain is still maturing and shows avenues for further research. Table~\ref{table: Sustainability in Blockchain} highlights the state-of-art solutions that achieve sustainability objectives within the blockchain. IPFS seems to be the most discussed solution within off-chain storage compared to StorJ, Sia, Swarm and Arweave. Research focusing on the later off-chain techniques can help further expand the applications within the blockchain. Furthermore, blockchain forks are an issue under rewritability as they introduce the possibility of altering the transaction history, and changing the state of the blockchain research around this can maximise the utilisation of the redactable solutions.

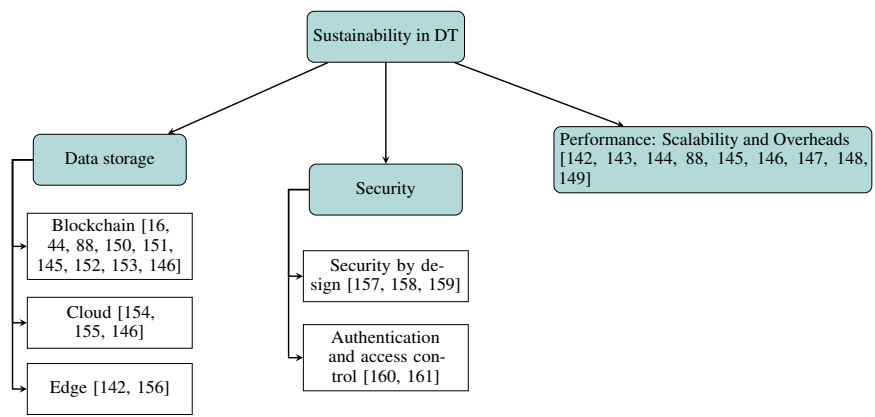
\begin{figure*}[!ht]
    \centering
    \resizebox{0.7\textwidth}{!}{
    \begin{tikzpicture}[node distance=2cm]    
    \tikzstyle{decision} = [rectangle, rounded corners, minimum width=3cm, minimum height=1cm,text centered, draw=black,fill=teal!30]
    \tikzstyle{rect} = [rectangle, minimum width=3cm, minimum height=1cm, text width=3cm, text centered, draw=black]
    \tikzstyle{arrow} = [thick,->,>=stealth]
    
    \node (start) [decision] {Sustainability in DT};
    \node (child1) [decision, below left of=start, xshift=-4cm, yshift=-1cm] {Data storage};
    \node (child2) [decision, below of=start, yshift=-1cm] {Security};
    \node (child3) [decision, below right of=start, xshift=5cm, yshift=-1cm] {{Performance}};
    \node (rect1) [rect, below of=child1, yshift=0.3cm] {Blockchain~\cite{SUHAIL10.1145/3517189, WANG9384115, PUTZ2021102425, LENG8789508, TAKAHASHI9754902, WANG9718531, YAQOOB9076112, KHAN9310354, SHEN2021338}};
    \node (rect2) [rect, below of=child1, yshift=-1.2cm] {Cloud~\cite{WANG2022124, LIU2022108488, SHEN2021338}};
    \node (rect3) [rect, below of=child1, yshift=-2.5cm] {Edge~\cite{ZHAO10073211, HUANG2021138}};
    \node (rect4) [rect, below of=child2, yshift=0.3cm] {Security by design~\cite{WANG10061664, HOLMES9566277, walsh_cybersecurity_2019}};
    \node (rect6) [rect, below of=child2, yshift=-1.3cm] {Authentication and access control~\cite{CHENhttps://doi.org/10.1002/ett.4751, LIU9810813}};
    \node (rect7) [rect, below of=child3, yshift=0.2cm] {Scalability and Overheads~\cite{ZHAO10073211, BARNI8710554, QI8740963, PUTZ2021102425, WANG9718531, SHEN2021338, CHAVEZ10015336, SEOK9864249, NGUYEN9353208}};
    \node (rect8) [rect, below of=child3, yshift=-1.7cm] {Modelling and Simulation~\cite{OKEGBILE10488084, SERRANORUIZ2024100582, CHEN2024118974, MUSTOFA2024100633, VASSILEV2024105419, KARKARIA2024322, STADTMANN2024125169}};
    
    \draw [arrow] (start) -- (child1);
    \draw [arrow] (start) -- (child2);
    \draw [arrow] (start) -- (child3);
    \draw [arrow] (child1.west) -- ++(-0.4cm,0) |- (rect1.west);
    \draw [arrow] (child1.west) -- ++(-0.4cm,0) |-  (rect2.west);
    \draw [arrow] (child1.west) -- ++(-0.4cm,0) |-  (rect3.west);
    \draw [arrow] (child2.west) -- ++(-0.4cm,0) |- (rect4.west);
    \draw [arrow] (child2.west) -- ++(-0.4cm,0) |-  (rect6.west);
    \draw [arrow] (child3.west) -- ++(-0.4cm,0) |- (rect7.west);
    \draw [arrow] (child3.west) -- ++(-0.4cm,0) |-  (rect8.west);
    
    \end{tikzpicture}
    }
    \caption{A broad classification of sustainability features in DT.}
    \label{fig: Classification of sustainability in DT}
\end{figure*}
\section{Taxonomy of Sustainability with DT}\label{sec: Section 5}

\begin{figure}[!h]
  \centering
  \includegraphics[width=0.9\linewidth]{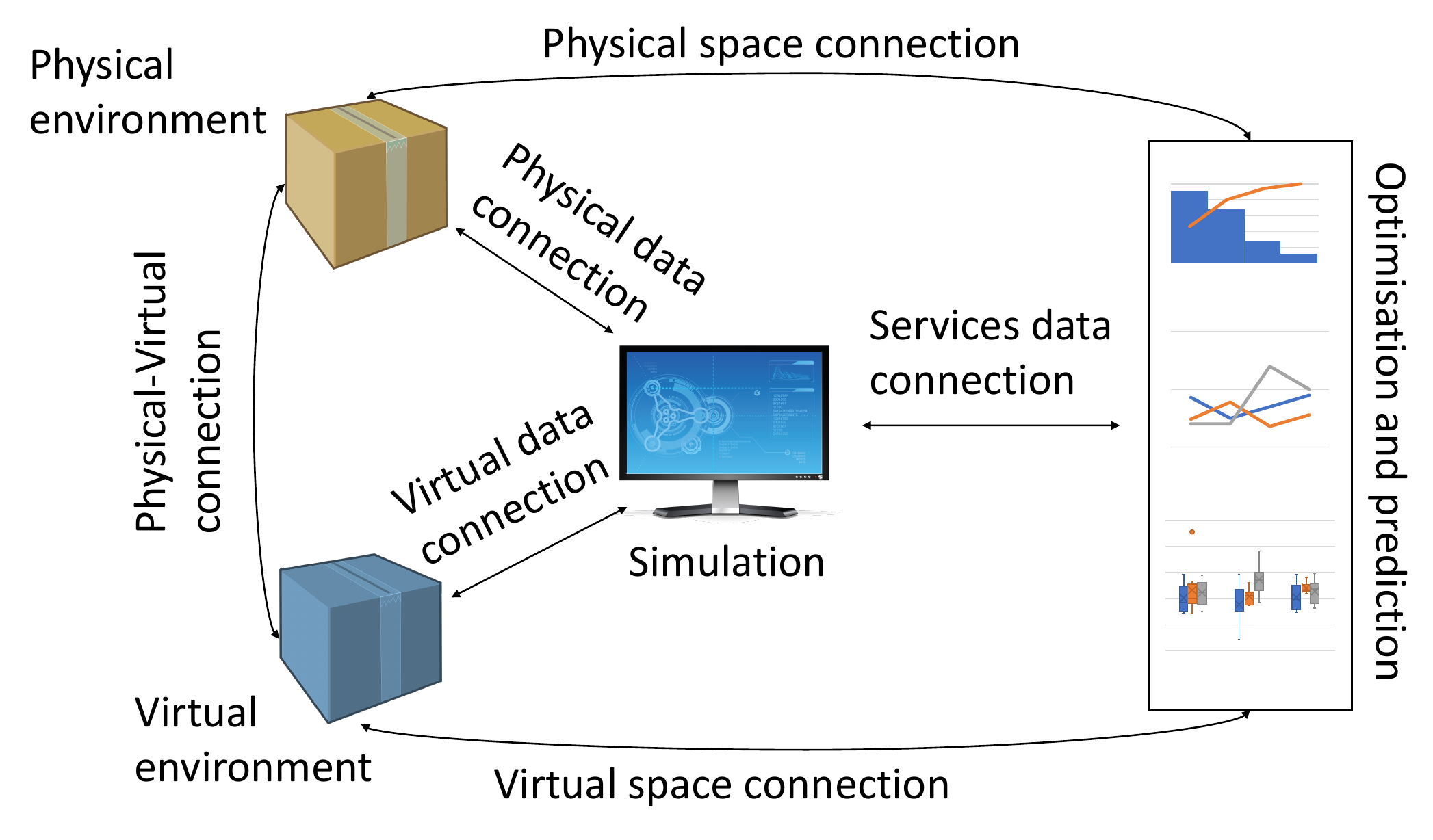}
  \caption{An exemplary illustration of the 5D DT model~\cite{TAO8477101}.}
  \label{fig: 5D model} 
\end{figure}

DTs aim at integrating the physical and virtual data of a product lifecycle, which is enabled by vast amounts of data to support complex analytics~\cite{TAO8477101}. A DT can follow different types of architecture though, for simplicity, a 5D version (as seen in Fig.~\ref{fig: 5D model}) is discussed. This includes: (1) Physical environment; (2) Virtual environment; (3) Data and connectivity; (4) Optimisation and prediction; and (5) Simulation~\cite{TAO2018169}. 

However, with the growing applications of DT, there is still no universally accepted or standardised set of practices for developing or deploying a DT~\cite{TAO8477101}. Somers \textit{et al.}~\cite{SOMERS2023107145} evaluated different applications that use DTs in a cyber-physical system, suggesting that most of its applications are in manufacturing, therefore highlighting the need for an oracle to standardise the use of DT practices. Since DT models are based on real-time and historical data, having a large data volume will help with accurate forecasting and optimisation; however, in current practices, sustainability is not achieved when storing data~\cite{ZHAO10073211}. In addition, ensuring that data is securely stored is another sustainability issue where accurate statistical models and reduced computation overheads can further improve the model's efficiency as elaborated through the taxonomies presented in Fig~\ref{fig: Classification of sustainability in DT}.
\subsection{Data storage}
Due to the nature of DT, there is a continuous flow of data between all architecture components (as seen in  Fig.~\ref{fig: 5D model}). This data needs to be stored efficiently to be accessed for forecasting. To handle data storage, blockchain, cloud and edge storage are some of the critical solutions available in the literature and discussed below.  
\subsubsection{Blockchain}
Smart contract automation, traceability and data security are achieved within a blockchain-based DT that enhances sustainability. Suhail \textit{et al.}~\cite{SUHAIL10.1145/3517189} have discussed blockchain-based DTs and evaluated their benefits. Yaqoob \textit{et al.}~\cite{YAQOOB9076112} have discussed the different taxonomies of DT, barriers and future challenges within DT and blockchain-based DT. For example, one of the barriers to adopting blockchain deals with slow throughput, which could impact the performance of a DT. However, research by Putz \textit{et al.}~\cite{PUTZ2021102425} handled this by using Swarm, which is an off-chain storage technique. 
\subsubsection{Cloud}
A cloud offers ample storage and powerful computing capabilities. A cloud network can scale and is easily accessible. Several studies have confirmed the feasibility of cloud computing in manufacturing companies \cite{LIU2022108488}. However, even cloud platforms suffer inefficiency and latency caused by the limited bandwidth of an industrial Internet. 
\subsubsection{Edge}
Processing and analysing the data on or near the edge device is called edge computing or DT-edge network. Edge storage provides proximity, localised processing, reduced latency and bandwidth optimisation. Huang \textit{et al.}~\cite{HUANG2021138} and Zhao \textit{et al.}~\cite{ZHAO10073211} have realised the need for edge computing to reduce data transmission latency and improve the performance of the DT.

\subsection{Security}
The importance of security within DT seems to be overthrown by its financial benefits \cite{HOLMES9566277}. DT is one of the prominent technologies in Industry 4.0, and neglecting cybersecurity may contribute to flawed decision-making~\cite{WANG10061664}. The research focusing on sustainability and security within DT is preliminary and presented in Table~\ref{table: Sustainability in DT}.

\subsubsection{Security-by-design}
Security by design includes the best practice of incorporating security considerations and measures throughout the lifecycle of a DT system. The most popular solution is using a distributed ledger technology such as blockchain~\cite{SUHAIL10.1145/3517189, WANG9384115, PUTZ2021102425, LENG8789508, TAKAHASHI9754902, WANG9718531, YAQOOB9076112, KHAN9310354, SHEN2021338}. With more robust security measures, system sustainability is achieved by assisting against security threats, preserving data integrity and privacy, ensuring system availability and reducing the need for extensive security remediation operations. Wang \textit{et al.}~\cite{WANG10061664} introduced a secure DT architecture that addresses computing and physical layer security. Pointer-based Advanced Encryption Standard (AES) encryption is utilised for computing security, while the Hybrid Automatic Repeat Request technology with Reed-Solomon-based Forward Error Correction coding is used for physical layer security to ensure bit error correction and data secrecy. The physical layer security technique avoids eavesdroppers by using the properties of wireless channels, such as noise, fading and interference.  
\subsubsection{Authentication and Access Control}
With authentication and access control, the data transferred between the systems can be verified, and sustainability is maintained through system integrity, promoting regulatory compliance, optimising resource utilisation and improving stakeholder confidence. Chen \textit{et al.}~\cite{CHENhttps://doi.org/10.1002/ett.4751} and Liu \textit{et al.}~\cite{LIU9810813} have introduced DTs in the Internet of Vehicles (IoV) technology. The IoV DT's communication channel requires authentication and access control through the public internet. Chen \textit{et al.} addressed this by using blockchain and a Group Authentication and Privacy-preserving (GAP) scheme, with robust security features such as efficient key management and non-tampering. 

\subsection{Performance}
A DT integrates sensor data, historical information, and simulations to facilitate monitoring, analysis, and performance optimisation for enhanced decision-making. However, challenges like high-performance modern hardware and software requirements, resources, uncertainty and environmental viability, organised and processed data, real-time data analysis, security concerns, and computational demands must be addressed~\cite{ESHAGHI2024107342}.

\subsubsection{Scalability and Overheads}
The ability to scale the DT system enhances the quality of results as it can handle large quantities of diverse and complex data~\cite{QI8740963}. As a result, scalability in DTs can be achieved through efficient data storage~\cite{QI8740963, PUTZ2021102425, SHEN2021338}. Nguyen \textit{et al.}~\cite{NGUYEN9353208} highlighted the limitations of current monitoring systems, specifically within the CyberPhysical Cloud Manufacturing environment (CPCM) and introduced cloud manufacturing based on DTs, which supports sustainability by improving efficiency. Computational resources and processing time to execute DTs may significantly cause some overheads. For example, Seok \textit{et al.}~\cite{SEOK9864249} have researched State-Class Graphs (SCG), which are used to help evaluate what-if scenarios. They test their hypothesis on Time-Coloured Petri Net (TCPN) DTs and identify high overheads. To further improve SCGs, an optimised version is presented in their article. Scalability and controlled computing overheads may help enhance performance optimisation by ensuring efficient resource utilisation and operational efficiency.

\subsubsection{Modelling and Simulation}
With the growing application of DT, a homogeneous path to handle computational demand has not been defined~\cite{ESHAGHI2024107342, SERRANORUIZ2024100582}. Much of the recent works have explored different techniques to help improve sustainability in DT. Serrano-Ruiz \textit{et al.}~\cite{SERRANORUIZ2024100582} have focused on the job shop scheduling problem with constant demand fluctuation and interruptions in the manufacturing process. To tackle this, they have proposed a three-step process- the schedule is first translated to a sequential decision-making problem based on Markov Decision Processes (MDP) rules. Deep Reinforcement Learning (DRL) is then used to derive a policy to address the problem based on a trained proximal policy optimisation algorithm, which then decides on the next steps based on the predefined rule set to be applied to the job queue, which helps improve the overall efficiency of the system. Okegbile \textit{et al.}~\cite{OKEGBILE10488084} have introduced the concept of human DT, which is an emerging area of research. The need for constant data sharing among physical and virtual twins is critical, particularly in zero-trust environments where security is compromised. Therefore, to tackle this problem, they have proposed a shard-based blockchain which enables data sharing. The blockchain follows a priority-based block appending technique based on queuing theory MDP and DRL to optimise throughput and overall system performance. Chen \textit{et al.}~\cite{CHEN2024118974} integrates the Bayesian model as a part of modelling the DT as it provides more accurate parameter estimation by accounting for uncertainty and updating the current state of the structure when new data becomes available. Based on the results, the approach has shown effectiveness and notable efficacy in protecting non-linear structural behaviour. Karkaria \textit{et al.}~\cite{KARKARIA2024322} have developed a Bayesian inference model that uses a long short-term memory architecture. Monte Carlo simulation is used to gauge predictive uncertainty. This implementation emphasises significant historical events while also quantifying the uncertainty of the model.

\subsection{Summary}
DTs play a critical role in manufacturing systems and contribute to maximising system efficiency. They are being widely adopted in current systems. This section has explored the benefits of DTs and identified taxonomies that define sustainability within DT. Table~\ref{table: Sustainability in DT} highlights the different aspects of sustainability and research contributing to improving DTs. It is evident that DTs can support many applications; however, the cybersecurity of DT is an essential area of research that needs further exploration, with an emphasis on data management and pushing it to the blockchain.

\begin{landscape}
\begin{table}[!h]
    \centering
    \resizebox{23.5cm}{!}{
    \begin{tabular}{|p{2cm}|p{5cm}|c|c|c|c|c|p{3cm}|c|p{5cm}|}
        \hline
        \multirow{2}{*}{\textbf{Author (Year)}} & \multirow{2}{*}{\centering\textbf{Ideology}} & \multicolumn{5}{c|}{\textbf{Sustainability features in DT}} & \multirow{2}{*}{\centering\textbf{Use case}} & \multirow{2}{*}{\textbf{Comparison}} & \multirow{2}{*}{\centering\textbf{Parameters}} \\
        \cline{3-7}
        & & \textbf{Fidelity} & \textbf{Throughput} & \textbf{Execution time} & \textbf{Access control} & \textbf{Backup and recovery} & & &\\
        \hline
        [Chen \textit{et al.} 2023]~\cite{CHENhttps://doi.org/10.1002/ett.4751} & Authorise DT data using blockchain-based GAP scheme & \centering - & \centering - & low & {\ding{51}} & \centering - & Internet of Vehicles & {\ding{51}} & \multicolumn{1}{c|}{\begin{tabular}[c]{@{}c@{}} Loss \\ Prediction accuracy \\ Encryption time \\ Negotiation time \\ Calculation time \end{tabular}} \\
        \hline
        [Huang \textit{et al.} 2021]~\cite{HUANG2021138} & Edge-based DT for high performance and low latency & \centering - & high & low & - & - & Real-time health monitoring of industrial systems and anomaly detection & - & \multicolumn{1}{c|}{\begin{tabular}[c]{@{}c@{}} -\end{tabular}} \\ 
        \hline
        [Khan \textit{et al.} 2022]~\cite{KHAN9310354} & Run Twinchain with DT to handle the limitations with blockchain via a 6D DT framework & - & - & - & - & - & Supply chain & - & \multicolumn{1}{c|}{\begin{tabular}[c]{@{}c@{}} - \end{tabular}} \\ 
        \hline
        [Leng \textit{et al.} 2020]~\cite{LENG8789508} & A bi-level intelligence technique where the blockchain is responsible for data organisation in one level and the DT handles optimisation in the second level & \centering - & high & low & {\ding{51}} & - & Industrial Internet of Things & {\ding{51}} & \multicolumn{1}{c|}{\begin{tabular}[c]{@{}c@{}} - \end{tabular}} \\ 
        \hline
        [Liu \textit{et al.} 2021]~\cite{LIU2022108488} & A four-terminal architecture for controlling thermal error of machines, four terminals are used to handle inefficiency and latency of cloud computing. & {\ding{51}} & high & medium & - & {\ding{51}} & Manufacturing plant & - & \multicolumn{1}{c|}{\begin{tabular}[c]{@{}c@{}} - \end{tabular}} \\ 
        \hline
        [Liu \textit{et al.} 2022]~\cite{LIU9810813} & A new privacy-preserving authentication technique based on BAN logic, ROR model and informal security analysis & {\ding{51}} & - & - & {\ding{51}} & - & Internet of Vehicles & {\ding{51}} & \multicolumn{1}{c|}{\begin{tabular}[c]{@{}c@{}} Running time \\ Computation cost \end{tabular}} \\ 
        \hline
        [Putz \textit{et al.} 2020]~\cite{PUTZ2021102425} & Ensure secure data sharing while maintaining availability, integrity and confidentiality & {\ding{51}} & high & low & {\ding{51}} & \centering - & Manufacturing & \centering - & \multicolumn{1}{c|}{\begin{tabular}[c]{@{}c@{}} Runtime \end{tabular}} \\ 
        \hline
        [Seok \textit{et al.} 2023]~\cite{SEOK9864249} & Test TCPN DTs to evaluate the performance of SCGs & {\ding{51}} & high & low & - & - & Manufacturing & {\ding{51}} & \multicolumn{1}{c|}{\begin{tabular}[c]{@{}c@{}} Execution time \end{tabular}} \\
        \hline
        [Shen \textit{et al.} 2021]~\cite{SHEN2021338} & A blockchain-based DT utilising the cloud as off-chain storage & {\ding{51}} & medium & medium & {\ding{51}} & \centering - & \centering - & {\ding{51}} & \multicolumn{1}{c|}{\begin{tabular}[c]{@{}c@{}} Latency \\ Optimal profit \\ Optimal sampling rate \end{tabular}} \\ 
        \hline
        [Wang \textit{et al.} 2021]~\cite{WANG9718531} & Optimise data fidelity and enhance energy and information sustainability & {\ding{51}} & high & low & {\ding{51}} & \centering - & \centering - & {\ding{51}} & \multicolumn{1}{c|}{\begin{tabular}[c]{@{}c@{}} Average reveal delay \\ Depletion rate in the simulation \\ Average data fidelity \\ No. of devices \end{tabular}} \\ 
        \hline
        [Wang \textit{et al.} 2021]~\cite{WANG2022124} & Cloud-based DT system for manufacturing systems & {\ding{51}} & high & low & {\ding{51}} & - & Battery manufacturing & {\ding{51}} & \multicolumn{1}{c|}{\begin{tabular}[c]{@{}c@{}} - \end{tabular}} \\ 
        \hline
        [Wang \textit{et al.} 2023]~\cite{WANG10061664} & A secure data exchange environment for a DT based on AES encryption and HARQ technique for physical layer encryption & {\ding{51}} & high & low & {\ding{51}} & \centering - & \centering - & {\ding{51}} & \multicolumn{1}{c|}{\begin{tabular}[c]{@{}c@{}} Energy consumption \\ Average number of transmissions \\ Energy consumption \\ Encrypted transmission delay \end{tabular}} \\ 
        \hline
        [Zhao \textit{et al.} 2022]~\cite{ZHAO10073211} & Using edge networks to enable low energy consumption and low computation for DT data & \centering - & high & low & \centering - & \centering - & 6G networks & {\ding{51}} & \multicolumn{1}{c|}{\begin{tabular}[c]{@{}c@{}} Energy consumption \end{tabular}} \\ 
        \hline
    \end{tabular}
    }
    \caption{State-of-the-art approaches applicable for sustainability in DT ({\ding{51}} Discussed and specified; - Not discussed and/or not specified).}
    \label{table: Sustainability in DT}
\end{table}
\end{landscape}

\section{Computing problems and research questions}\label{sec: Section 6}

%\subsection{Computing Problems and research trends}
Blockchain and DT exhibit characteristics that can enable sustainability within BDM. However, to realise the benefits of these two technologies, the following areas need to be addressed: 

\begin{figure*}[!ht]
  \centering
  \includegraphics[width=1\textwidth]{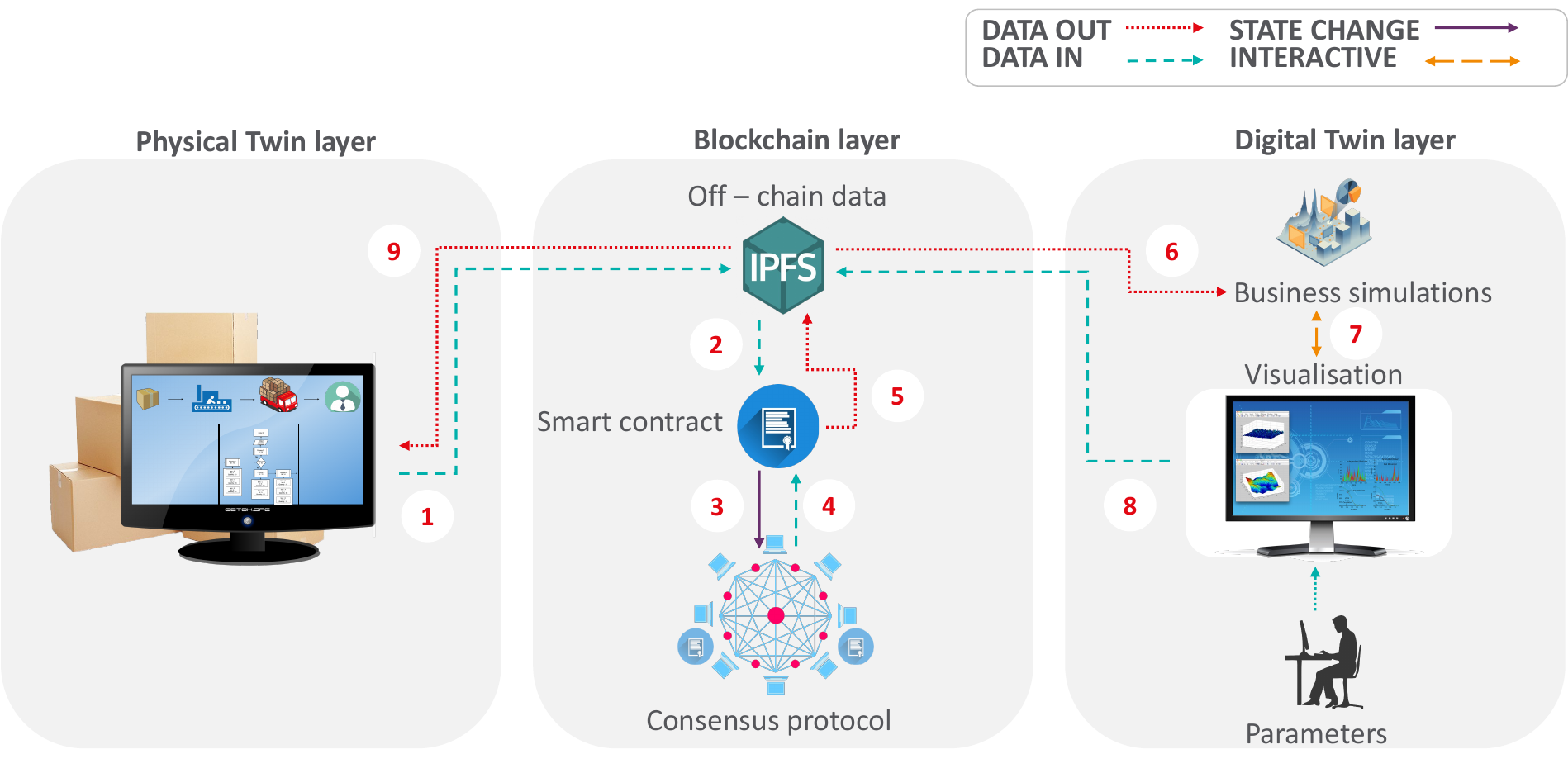}
  \caption{An illustration of the blockchain's workflow for a supply chain.}
  \label{fig: Workflow processes} 
\end{figure*}

\subsection{Blockchain} 
\begin{itemize}
    \item Security and Resilience - The research around security features in blockchain has been primarily discussed through implementations~\cite{ZHANG10.1145/3316481}. However, an increasing interest in quantum resilience within blockchains is a growing area of research, and further efforts can focus on the sustainable implementation towards quantum resilience~\cite{DOLEV10041785, SUHAIL10.1145/3517189}. Another security concern within blockchains arises due to rewritability~\cite{YE10004754}, where research can focus on techniques to strengthen security using non-cryptographic techniques. Furthermore, within cryptographic rewritable techniques, key revocation and access control monitoring are areas of research that require further evaluation. Consensus protocols have had a plethora of research contributions \cite{MILUTINOVIC10.1145/3007788.3007790}; however, ensuring optimum security and performance within public blockchains still needs a ready-to-deploy solution \cite{kokoris-kogias_enhancing_nodate}.  
    \item Scalability - To ensure efficient scalability of blockchains without much latency, further research can focus on off-chain data storage, specifically in data retrieval and storage. Although sharding aims to improve scalability~\cite{LI10026494}, the cross-sharding issue affects data security and authenticity, which would require further research. Additionally, it is difficult for sidechains to work efficiently \cite{YIN9543599} if blockchains are scaled with branching. 
    % \item Privacy and confidentiality - \cite{YE10004754}
    \item Social and Economic Implications - This includes studying the impact on industries, socio-economic inequalities and digital inclusion, as well as examining the potential for new business models in how they can contribute to a circular economy and sustainable development~\cite{ESMAEILIAN2020105064}. 
    \item Governance - Zhang \textit{et al.}~\cite{ZHANG2020104512} have discussed the importance of LCA, which impacts a product lifecycle's environmental and social footprint. LCA benefits from blockchain through its complexities in tracking and quantifying inputs and outputs at multiple supply chain stages. However, Key Performance Indicators (KPIs) are an underlying contributor towards LCA. Research exploring KPIs within the context of Industry 4.0 has not been significantly standardised. Furthermore, future research could identify the barriers to blockchain-based BDM approaches and explore ways to tackle these issues, enabling a structured approach to follow industry standards.
\end{itemize}

% not considered backup and recovery for data storage.
\subsection{DT}
\begin{itemize}
    \item Security - If a DT is compromised, it risks exposing organisational information or may lead to further back-end system attacks, as these systems may be called directly by the twin~\cite{HOLMES9566277, CHEN.2236}. Therefore, research covering security standards and best practices can help standardise DTs across industries. 
    \item Resilience - DTs are currently implemented differently across industries. There is no standardised approach that is practised~\cite{HUANG2021138}. Establishing a standard approach requires further research to identify the most suitable architecture. 
    \item Access control - Access rights are not distributed for data manipulation within a DT system~\cite{KHAN9310354, LENG8789508}. As a result, this could affect the prediction and optimisation models within BDM. Research on access control could ensure further security and tamper-proof DTs within industries.
\end{itemize}

\subsection{Open Questions}
This section highlights the different research questions based on the gaps in the literature. The details are:
\begin{itemize}
    \item RQ 1 - It is necessary to fully understand the goals and objectives of sustainability to apply the factors to this research - Here, the emphasis must be given to the criteria used to define the blockchain and DT policies. Such criteria can help with scores such as ESG. With traceable features of blockchain, a proven record of correct business practices can be managed effectively.   
    \item RQ 2 - BDM is a term not often expressed via mathematical modelling for broader applications, rather driven by policies which are harder to implement in the real world - It is crucial to define the purpose of BDM  to apply it within different research areas. Here, the emphasis must be on deriving mathematical reasoning within the decision components to make them applicable to broader use cases. 
    \item RQ 3 - Categorisation of sustainability vs performance needs to be elaborated, and their attainment needs more physical meaning from the computing perspective - Here, systems must be checked for their performance not just in terms of operational efficiency but for their impact on different factors of sustainability. Moreover, the tradeoffs need to be considered carefully, and any dilemma associated with it must be handled depending on the critical nature of the applications. 
    \item RQ 4 - The impact of legacy libraries in developing blockchain and DT applications can severely impact the sustainability of the overall system. Here, the emphasis needs to be on the technology that is used for developing the applications driven by underlying blockchain layer particularly their consensus mechanisms as these can severely impact the performance, Moreover, using libraries that consume more computing cycles and are not energy efficient can negatively impact BDM and associated sutaibaility with each iteration. 
    \item RQ 5 - Handling heterogeneous data in DT requires several integrated solutions, and their optimisation can severely impact the sustainability, which needs to be accounted for in their modelling - Here, the focus should be towards sustainable practices towards prediction and forecasting features of DT, which can consume much of the resources and can affect the ESG scores of the systems relying on DT for their decision making. Emphasis can be given to simple methods like building solutions to avoid retraining the models on the same data. 
\end{itemize}

\section{Case study}\label{sec: Section 7}

This survey uses a case study of a supply chain to show the impact of sustainability in BDM with blockchain and DT. Specifically, we consider the Bill of Material (BOM) case as it covers supplier management, quality control, cost estimation and maintenance \cite{BARCLAY2019}. 

\subsection{Architecture design}
Fig. \ref{fig: Workflow processes} defines the different steps within the integrated blockchain and DT-based BDM process. Smart Contracts are used within the blockchain to interact and establish a connection between the supply chain and the DT. The system architecture is built on three layers:
\begin{itemize}
  \item Physical layer - The physical layer consists of the real-time supply chain, which feeds data to the analytics engine. Data that is extracted by the application includes information such as Part ID, Part Name, Description, Procurement Type, Comments and Notes, Quantity and Priority. This data is processed through data simulation and modelling, which is an integral part of BDM. 
  \item Blockchain layer - The blockchain layer acts as the decentralised database management of the BDM. The blockchain communicates with the Physical and DT layers using smart contracts. To improve the scalability of the blockchain, an off-chain database is utilised. For example, data is transferred from the physical layer to IPFS, which returns a hash value \cite{IPFS_docs_2021}. The hash value and other meta-data, including information such as Sender, Recipient, Amount, Signature and Employee Id, produce a new transaction. Several transactions create a block to be added to the blockchain. The decentralised nature of the blockchain is managed by broadcasting blocks between active nodes running a consensus protocol. Table \ref{table: Consensus comparison} gives a comparison of private blockchains that are applicable for supply chain applications \cite{BELCHIOR10.1145/3471140, MOHAN10.1145/3299869.3314116}. A number of research papers \cite{HUANG10.1145/3441692, ISLAM9860157, XIAO8972381, SAMUEL9461085} discuss consensus protocols in more detail.
  \item DT layer - The DT layer is a digital replica of the real-time supply chain. Data is fed to the DT by the blockchain through a smart contract. The DT consists of optimisation and prediction models that simulate different scenarios. Scenarios are provided in terms of parameters by a user. For example, the prediction models can identify the possible effects on the input parameters and propose the next best alternative through the optimisation model to ensure that the supply chain is unaffected.
\end{itemize}

\begin{table*}[htbp]
    \centering
    \resizebox{18cm}{!}{
    \begin{tabular}{|p{1.5cm}|p{3cm}|c|p{1.8cm}|p{1.8cm}|p{1.5cm}|p{4cm}|p{4cm}|}
        \hline
        Consensus protocol & Blockchain platform & Energy efficiency & Complexity-throughput & Transaction cost & Forks & Problem addressed & Limitation\\
        \hline
        \centering PoA \cite{ISLAM9860157, WANG8629877} & \centering Aura, Clique & \raisebox{-1\height}{\begin{tikzpicture} \node[rectangle, rounded corners, shading=axis, left color=black, middle color=black, right color=black, minimum width=3cm, minimum height=1cm] { }; \end{tikzpicture}} & \centering $O(n)$ & \centering {\ding{55}} & \centering {\ding{51}} & Replaces PoW and PoS in the permissioned environment & Based on the implementation of PoA the chance of forks are high, and it is not scalable\\
        \hline
        \centering Raft \cite{WANG8629877, XU10.1145/3579845} & \centering Apache Zookeeper & \raisebox{-0.6\height}{\begin{tikzpicture} \node[rectangle, rounded corners, shading=axis, left color=gray!50, middle color=gray!50, right color=white, minimum width=3cm, minimum height=1cm] { }; \end{tikzpicture}} & \centering $O(n^2)$ & \centering {\ding{51}} & \centering - & Handles Crash Fault Tolerant nodes & Unable to handle Byzantine Fault Tolerant nodes \\
        \hline
        \centering QBFT \cite{ABDELLA9344670, FAHMIDEH10.1145/3530813} & \centering Hyperledger Besu & \raisebox{-0.8\height}{\begin{tikzpicture} \node[rectangle, rounded corners, shading=axis, left color=black, middle color=white, right color=white, minimum width=3cm, minimum height=1cm] { }; \end{tikzpicture}} & \centering $O(n^2)$ & \centering {\ding{55}} & \centering {\ding{55}} & Improved efficiency over PBFT & Complexity is not linear, therefore it affects scalability\\
        \hline
        \centering PBFT \cite{HUANG10.1145/3441692, CACHIN2017blockchain} & \centering Hyperledger Fabric & \raisebox{-0.8\height}{\begin{tikzpicture} \node[rectangle, rounded corners, shading=axis, left color=black, middle color=white, right color=white, minimum width=3cm, minimum height=1cm] { }; \end{tikzpicture}} & \centering $O(n^2)$ & \centering {\ding{55}} & \centering {\ding{55}} & Built to handle byzantine fault tolerant nodes without any need of miners & Not scalable and complexity is quadratic\\
        \hline
        \centering PoET \cite{BASHAR9014382, XU10.1145/3579845} & \centering Hyperledger Sawtooth & \raisebox{-0.8\height}{\begin{tikzpicture} \node[rectangle, rounded corners, shading=axis, left color=gray!50, middle color=gray!50, right color=white, minimum width=3cm, minimum height=1cm] { }; \end{tikzpicture}} & \centering $O(n^2)$ & \centering {\ding{55}} & \centering {\ding{55}} & Proposed by Intel to replace PoW by using an arbitrary time lag for each node & Uses third-party hardware - security threat as hardware can be exploited\\
        \hline
        \centering Tangle \cite{XU10.1145/3579845, FAN9129732} & \centering IOTA & \raisebox{-1\height}{\begin{tikzpicture} \node[rectangle, rounded corners, shading=axis, left color=gray!50, middle color=gray!50, right color=white, minimum width=3cm, minimum height=1cm] { }; \end{tikzpicture}} & \centering $O(n^2)$ & \centering {\ding{51}} & \centering {\ding{55}} & Uses Nakamoto consensus and implements DAG structure to solve scalability issues & Consists of fees and miners\\
        \hline
        \centering - \cite{MOHAN10.1145/3299869.3314116, FAHMIDEH10.1145/3530813} & \centering MultiChain & \raisebox{-0.8\height}{\begin{tikzpicture} \node[rectangle, rounded corners, shading=axis, left color=gray!50, middle color=gray!50, right color=white, minimum width=3cm, minimum height=1cm] { }; \end{tikzpicture}} & \centering $O(n)$ & \centering {\ding{55}} & \centering {\ding{51}} & An API that allows connection between different blockchains & Chance of forks are high\\
        \hline
    \end{tabular}
    }
    \caption{A comparison of private consensus protocols.}
    \label{table: Consensus comparison}
    \footnotesize
    \begin{flushleft}
      {\ding{55}} Does not exists; {\ding{51}} Exists; - Not mentioned; \begin{tikzpicture} \node[rectangle, shading=axis, left color=black, middle color=black, right color=black, minimum width=0.5cm, minimum height=0.22cm] { }; \end{tikzpicture} High efficiency; \begin{tikzpicture} \node[rectangle, shading=axis, left color=black, middle color=white, right color=white, minimum width=0.5cm, minimum height=0.22cm] { }; \end{tikzpicture} Medium efficiency;  \begin{tikzpicture} \node[rectangle, shading=axis, left color=gray!50, middle color=gray!50, right color=white, minimum width=0.5cm, minimum height=0.22cm] { }; \end{tikzpicture} Low efficiency
    \end{flushleft}
\end{table*}

\subsection{Performance evaluation}
\begin{figure}[htbp]
  \centering
  \begin{subfigure}{0.5\textwidth}
    \centering
    \includegraphics[height=180px, width=1\linewidth]{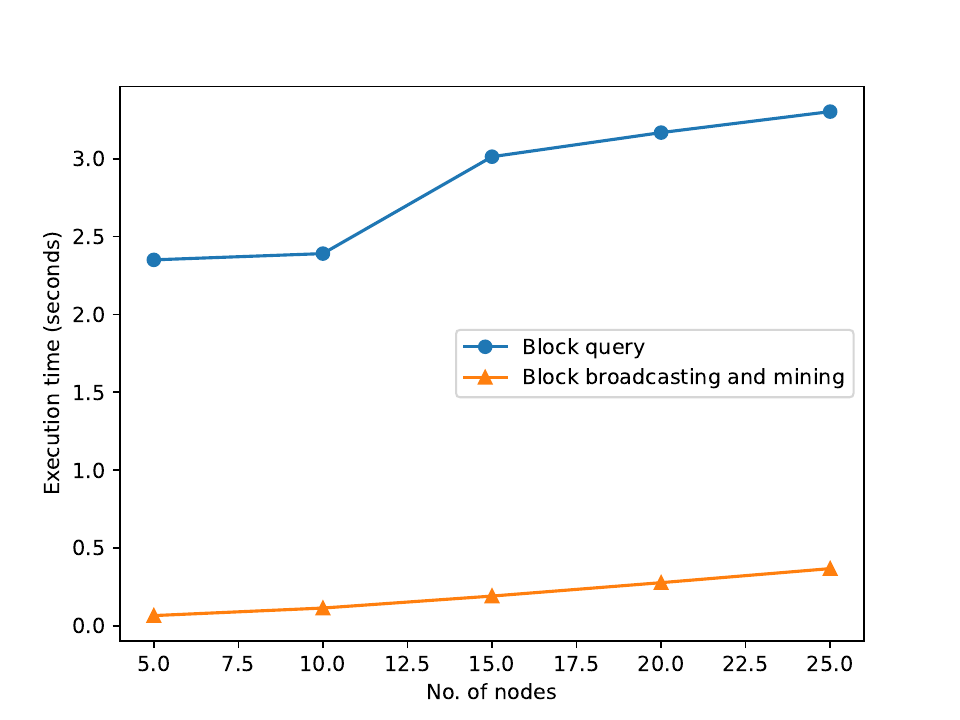}
    \caption{Execution time.}
    \label{fig: Execution time}
  \end{subfigure}
  \hfill
  \begin{subfigure}{0.45\textwidth}
    \centering
    \includegraphics[height=150px, width=1\linewidth]{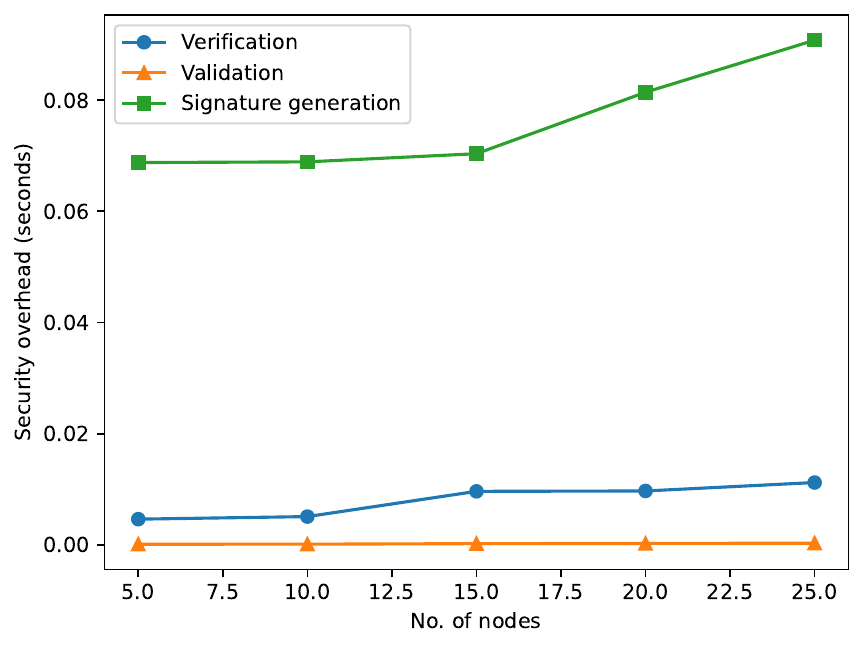}
    \caption{Security overheads.}
    \label{fig: Security overheads}
  \end{subfigure}
  \caption{Blockchain-based BOM results.}
  \label{fig: Blockchain-based BOM results}
\end{figure}
This subsection considers the advantages of a blockchain-based model towards sustainability. The model is executed in Python Flask, and the system connects to IPFS using Infura \footnote{Infura, \url{https://www.infura.io}}. The application is tested on a 2020 Mac Book Pro (model number - MYDC2B/A) running Mac-OS. The processor is an Apple M1 chip with eight cores @3.2GHz with 8GB RAM. Nodes were simulated using the Command Line Interface, where nodes ranged from 5 - 25. Experimental results were based on an average value obtained through ten runs. The consensus protocol used is a Proof of Work protocol. The difficulty target of the algorithm was set to finding a hash that begins with two zeros due to the system's limitations. The tests targeted execution time and security overhead (refer to Fig. \ref{fig: Blockchain-based BOM results}), which are two essential sustainability factors. The details are as follows:
\begin{itemize}
    \item Execution time - This contributes to energy efficiency, scalability and resource utilisation \cite{KLEINKNECHT9591261}. Based on the results in Fig. \ref{fig: Execution time}, the execution time for block broadcasting and mining is relatively consistent. However, there is an increase in block query time, which can be due to network latency issues as this application uses IPFS as an off-chain database.
    \item Security overheads - This enables trust, prevents financial losses and maintains regulatory compliance, which strengthens sustainability in BDM \cite{ZHANG10.1145/3316481, KLEINKNECHT9591261}. Following the results in Fig. \ref{fig: Security overheads}, the security overheads (in milliseconds) - although low and acceptable in current settings - would need considerable reduction for large-scale deployment to improve efficiency.
\end{itemize}
The experiments conducted had limitations such as network latency, hardware limitations and consensus algorithm variation, which could suggest further improvements to the application. The system has only been tested in a stand-alone environment and has not been deployed in a production environment.

\section{Methodology}\label{sec: Section 8}

The methodology section presents the outline of the selection process to determine the articles used for this research to answer the questions presented in Section \ref{sec: Section 6}. The research process involved the identification of relevant papers across a range of databases, including ACM Digital Library, IEEE Xplore, Web of Science and Scopus. Identifying primary research refined it further by applying particular inclusion criteria and synthesising the data through keyword matching and reading abstracts. Only the studies written in the English language have been included. A thorough literature survey was conducted once the relevant papers were identified. 

\begin{figure}
  \centering
  \includegraphics[width=0.5\textwidth]{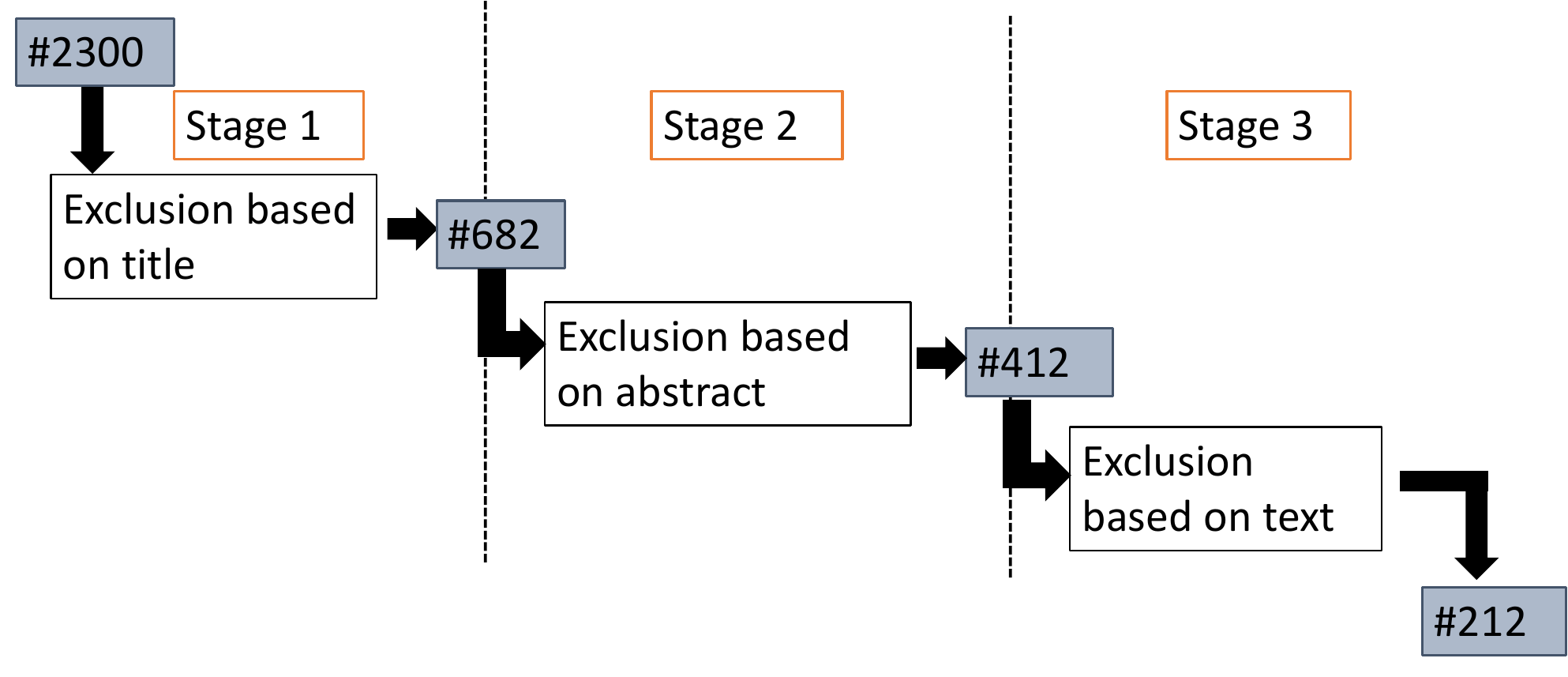}
  \caption{Study selection procedure.}
  \label{fig: methodology} 
\end{figure}

\begin{figure}
  \centering
  \includegraphics[width=0.5\textwidth]{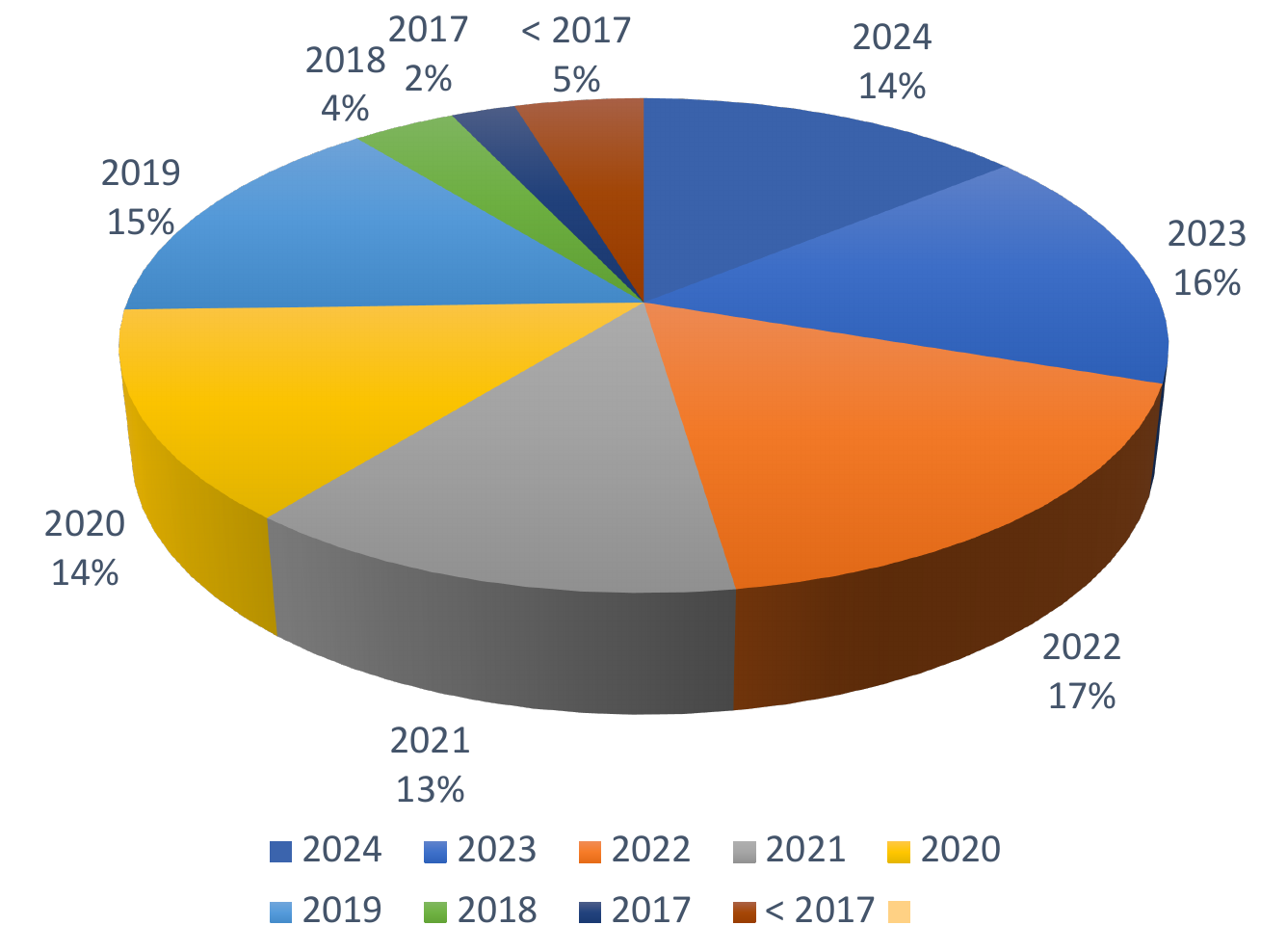}
  \caption{Paper distribution by year of publication.}
  \label{fig: distribution}
\end{figure}
\begin{figure}
  \centering
  \includegraphics[width=0.5\textwidth]{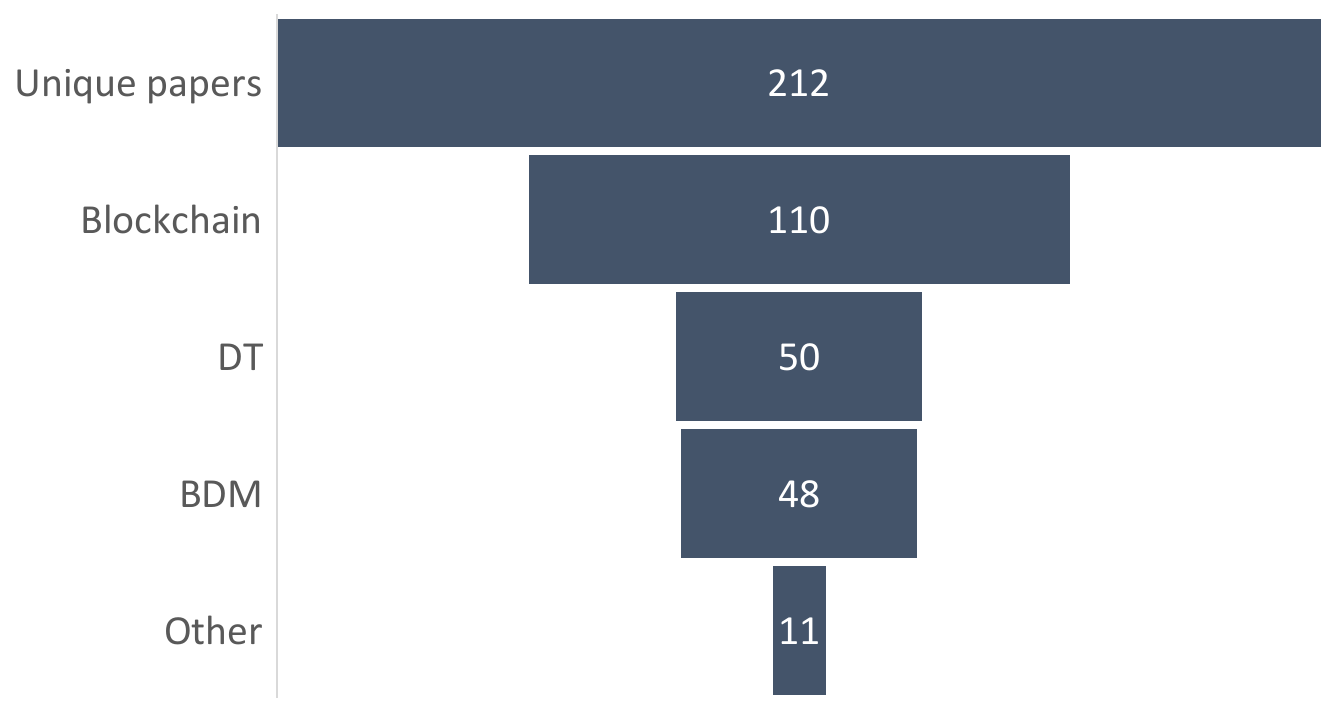}
  \caption{Topic distribution including overlap of disciplines.}
  \label{fig: topic distribution}
\end{figure}

The study selection procedure is shown in Fig. \ref{fig: methodology}. Many known research papers were excluded because their titles did not align with the criteria or the abstract was not close to the ideology of this survey. As shown in Fig. \ref{fig: methodology}, the initial search returned over 2202 papers, which were narrowed down to 621 papers based on their titles and 348 papers based on their abstracts. These 348 papers were further filtered to the final list of 170 papers. Fig. \ref{fig: distribution} presents this as a pie chart, and Fig. \ref{fig: topic distribution} shows how the research articles are distributed across the areas of Blockchain, DT and BDM.

\section{Conclusion}\label{sec: Section 9}
With the globalisation of industries, the need for transparency, access control and trusted data management is integral in ensuring sustainability in Business Decision Modelling (BDM). Effectively utilising these large abundances of data to improve industrial processes is fundamental to achieving efficiency in Industry 4.0. This survey covers an in-depth, comprehensive review of the existing state-of-art contributions towards BDM, blockchain and Digital Twin (DT). The key takeaway from this survey is that sustainability is essential as it allows for the efficient use of resources, focusing on economic, social and environmental features within industries. Here, BDM plays a vital role in handling risks and disruptions through predictive analysis. Sustainability in BDM focuses on the triple bottom line approach, specifically on social, economic and environmental factors. These can be evaluated using decision models based on prediction and optimisation. Blockchain can support sustainability through its decentralised architecture, providing transparency, security and data integrity. DT can support sustainability through resource optimisation, predictive maintenance, stakeholder collaboration and process optimisation. Furthermore, blockchain can help BDM improve data coordination, and DT can offer efficient data simulation processes and workflow management to accelerate sustainability.

 \bibliographystyle{elsarticle-num}
 \bibliography{references}

\end{document}